\documentclass[fleqn,usenatbib]{mnras}

\usepackage{newtxtext,newtxmath}

\usepackage[T1]{fontenc}

\DeclareRobustCommand{\VAN}[3]{#2}
\let\VANthebibliography\thebibliography
\def\thebibliography{\DeclareRobustCommand{\VAN}[3]{##3}\VANthebibliography}

\usepackage{graphicx}	
\usepackage{amsmath}	
\usepackage[printonlyused]{acronym}
\usepackage{soul}

\newcommand{\Msun}{\mbox{M$_{\odot}$}}
\newcommand{\Rsun}{\mbox{R$_{\odot}$}}

\newcommand{\M}[1]{\mbox{$M_\mathrm{#1}$}}

\newcommand{\parsec}{\mbox{\textsc{parsec}}}
\newcommand{\mesa}{\mbox{\textsc{mesa}}}

\newcommand{\amlt}{\mbox{$\alpha_\mathrm{MLT}$}}

\newcommand{\Gav}{$\langle \Gamma_1\rangle $}
\newcommand{\Tc}{\mbox{$T_\mathrm{c}$}}
\newcommand{\rhoc}{\mbox{$\rho_\mathrm{c}$}}

\acrodef{ZAMS}{zero age main sequence}
\acrodef{MS}{main sequence}
\acrodef{CHeB}{core helium burning}
\acrodef{COB}{core oxygen burning}
\acrodef{CC}{core collapse}
\acrodef{RSG}{red super giant}
\acrodef{BSG}{blue super giant}
\acrodef{HR}{Hertzsprung-Russell}
\acrodef{BH}{black hole}
\acrodef{PI}{pair-instability} 
\acrodef{SPH}{smoothed particle hydrodynamics}

\title[Evolution of a post-collision star]{Formation of black holes in the pair-instability mass gap: evolution of a post-collision star}

\author[G. Costa et al.]{
Guglielmo Costa,$^{1,2,3}$\thanks{E-mail: guglielmo.costa@unipd.it}
Alessandro Ballone,$^{1,2,3}$
Michela Mapelli,$^{1,2,3}$\thanks{E-mail:michela.mapelli@unipd.it}
Alessandro Bressan,$^{4}$
\\
$^{1}$Physics and Astronomy Department Galileo Galilei, University of Padova,  
Vicolo dell'Osservatorio 3,
I--35122, Padova, Italy\\
$^{2}$INFN - Padova,
Via Marzolo 8,
I--35131, Padova, Italy\\
$^{3}$INAF - Osservatorio Astronomico di Padova, Vicolo dell'Osservatorio 5,
I-35122, Padova, Italy\\
$^{4}$SISSA, via Bonomea 365, I--34136 Trieste, Italy
}

\date{Accepted XXX. Received YYY; in original form ZZZ}

\pubyear{2022}

\begin{document}
\label{firstpage}
\pagerange{\pageref{firstpage}--\pageref{lastpage}}
\maketitle

\begin{abstract}
The detection of GW190521 by the LIGO--Virgo collaboration has revealed the existence of black holes (BHs) in the pair-instability (PI) mass gap. 
Here, we investigate the formation of BHs in the PI mass gap via star -- star collisions in young stellar clusters. 
To avoid PI, the stellar-collision product must have a relatively small core and a massive envelope.
We generate our initial conditions from the outputs of a hydro-dynamical simulation of the collision between a core helium burning star ($\sim 58$~\Msun) and a main-sequence  star ($\sim 42$~\Msun). The hydro-dynamical simulation allows us to take into account the mass lost during the collision ($\sim 12$~\Msun) and to build the chemical composition profile of the post-collision star. 
We then evolve the collision product with the stellar evolution codes {\sc parsec} and {\sc mesa}.  
We find that the post-collision star evolves through all the stellar burning phases until core collapse, avoiding PI. At the onset of core collapse, the post-collision product is a blue super-giant star. 
We estimate a total mass loss of about 1~\Msun\ during the post-collision evolution, due to stellar winds and shocks induced by neutrino emission in a failed supernova. The final BH mass is $\approx{87}$ \Msun. Therefore, we confirm that the collision scenario is a suitable formation channel to populate the PI mass gap.
\end{abstract}




\section{Introduction}
\label{sec:intro}

Stellar collisions can lead to the formation of exotic stars, such as blue stragglers \citep[e.g.,][]{sigurdsson1994,sills1997,sills2001,mapelli2004,mapelli2006,glebbeek2008,ferraro2012,portegieszwart2018} and very massive stars \citep[e.g.,][]{portegieszwart2002,freitag2006,mapelli2016,boekholt2018}. In star clusters, stellar collisions are triggered by dynamical interactions, which perturb the orbit of binary stars \citep[e.g.,][]{portegieszwart1997,portegieszwart1999}. Multiple collisions among massive stars in a dense star cluster could also lead to the formation of an intermediate-mass \ac{BH} with mass $>100$ M$_\odot$ \citep[e.g.,][]{portegieszwart2004,guerkan2006,giersz2015,mapelli2016,dicarlo2021,rizzuto2021,torniamenti2022}.  This process is expected to occur in the early evolution of a star cluster, when the most massive stars sink to the center of the cluster by dynamical friction, undergoing multiple collisions before they die by core collapse \citep{portegieszwart2004}. Stellar metallicity is a key ingredient driving the formation of a massive \ac{BH} via stellar mergers, because stellar winds can dramatically suppress the formation of a very massive star in a metal-rich star cluster \citep[e.g.,][]{glebbeek2009,mapelli2016}.

\cite{dicarlo2019} proposed that stellar collisions in dense star clusters can also trigger the formation of \ac{BH}s in the \ac{PI} mass gap ($\sim{}60-120$ M$_\odot$). Massive metal-poor stars are expected to undergo \ac{PI} at the end of carbon burning, when their hydrostatic equilibrium is compromised by an efficient production of electron--positron pairs in the core. This generally happens if the star develops a He core $\gtrsim{}32$ M$_\odot$ and is thought to suppress the formation of \ac{BH}s with mass $\sim{}60-120$ M$_\odot$ (e.g., \citealt{heger2002,belczynski2016pair,woosley2017,woosley2019,spera2017,stevenson2019}, but see \citealt{farmer2019,farmer2020,mapelli2020,marchant2021,costa2021,farrell2021,vink2021} for possible uncertainties on the mass gap). By running direct N-body simulations coupled with population synthesis, \cite{dicarlo2019} found that the collision between an evolved star, with an already developed He core, and a \ac{MS} star can result in an exotic star with an oversized envelope with respect to the core. Such a massive star might avoid pair instability and collapse to a \ac{BH} with mass inside the gap \citep[see also][]{spera2019,kremer2020,dicarlo2020a,dicarlo2020b,renzo2020}. This new model is one of the most promising scenarios for explaining the formation of GW190521 \citep{abbottGW190521,abbottGW190521astro} and other gravitational wave event candidates \citep{abbottGWTC2.1,abbottGWTC3}, which are likely associated with \ac{BH}s in the pair instability mass gap.
\citet{Vigna2019} showed that stellar mergers are also a promising scenario to produce some
hydrogen-rich pulsational pair instability supernovae.

The main limitation of \cite{dicarlo2019} and similar simulations is that no mass loss is assumed during the collision: the merger product is modeled by assuming that the entire mass of the \ac{MS} star wraps around the evolved star. In a companion paper \citep{Ballone2022}
, we take the pre-collision properties of two massive stars from  \cite{dicarlo2020b}  and simulate their collision with a hydrodynamical code \citep[\textsc{StarSmasher};][]{gaburov2018}.  We find that $\sim{}12\%$ of the total mass is lost during the collision and that the merger product shows evidence of partial mixing between the MS and the evolved star.

Here, we take the outputs of the hydrodynamical simulation performed by \citeauthor{Ballone2022}
and study the evolution of the merger product with the stellar evolution codes \parsec\ \citep{bressan2012, costa2019a} and \mesa\ \citep[][]{Paxton2011, Paxton2013, Paxton2015, Paxton2018, Paxton2019}. The hydrodynamical simulation ends when the collision product has reached hydro-static equilibrium but is still far from thermal equilibrium. Here, we relax and evolve the collision product to the end of \ac{COB} with the \parsec\ code, and to the final collapse with the \mesa\ code. 

This article is organized as follows. In Sect.~\ref{subsec:methods}, we present the physics adopted in the \parsec\ and \mesa\ stellar evolutionary codes, and  describe the post-collision model. 
We show the results in Sect.~\ref{sec:Results}, and discuss them in Sect.~\ref{sec:Discussion}. In Sect.~\ref{subsec:Conclusions}, we draw our conclusions.

\section{Methods} 
\label{subsec:methods}
To compute the stellar model before and after the collision, we used the \parsec\ V2.0 stellar evolutionary code \citep{bressan2012, costa2019a, Costa2019b} and the \mesa\ code version 12778 \citep{Paxton2011, Paxton2013, Paxton2015, Paxton2018, Paxton2019}. 

\subsection{{\sc parsec} configuration and pre-collision models}
\label{sec:parsec}

For the \parsec\ stellar tracks, we adopt the \citet{Caffau2011} solar composition and compute non-rotating models with an initial metal content $Z = 0.0002$ and a helium content $Y = 0.249$ \citep[from solar calibrations by][]{bressan2012}.
We use the Schwarzschild criterion \citep{Schwarzschild1958} to define convective unstable borders. We use the mixing length theory framework to compute the diffusion coefficients within unstable regions, with \amlt~=~1.74. For the overshooting region, we use the penetrative overshooting scheme described in \cite{Maeder1975} and \cite{Bressan1981}. We use an   overshooting parameter\footnote{The overshooting parameter $\lambda_{\rm ov}$ is the mean free path in pressure scale height unity that can be traveled by a convective eddy before dissolving, across the convective border.} $\lambda_{\rm ov} = 0.4$ \citep{costa2019a}. We adopted the opacities, neutrinos, equation of state, nuclear reaction network, and stellar winds described in detail in \citet{costa2021}.

Following \citet{dicarlo2020b}, the two components of the binary system are a \ac{CHeB} star with an initial mass \M1~=~57.6~\Msun, and a \ac{MS} star with an initial mass of \M2~=~41.9~\Msun. In the simulation by \citet{dicarlo2020b}, the two stars merge after 4.3~Myr. Due to the intrinsic differences 
between the stellar tracks of \citet{dicarlo2020b} \citep[i.e. \textsc{mobse} evolutionary tracks,][]{Giacobbo2018} 
and \parsec\,, we decided to build the pre-collision primary star by matching the evolutionary phase of the two stars rather than the exact age of the collision.
Therefore, we computed the evolution of the primary star from zero-age MS (ZAMS) to the \ac{CHeB} phase, stopping the computation when the star has a central helium content $X_{\rm c} \mathrm{(He)} \sim 0.72$ (in mass fraction).
We evolve the secondary star until the central hydrogen is $X_{\rm c}\mathrm{(H)} \sim 0.5$, during the \ac{MS} phase.

At these selected stellar stages, the primary star has a helium core of \M{He}~=~27.16~\Msun, while the secondary star is in the middle of the \ac{MS} and has a convective core of 23.7~\Msun. It contains 25.8~\Msun\ of H and 16~\Msun\ of He, distributed between the core and the envelope. 
Figure~\ref{fig:profile} shows the chemical profiles of the primary and secondary \parsec\ models. 

The two stellar models are used as initial conditions for the SPH simulation by \citet[][briefly described in Section~\ref{sec:hydro_sim}]{Ballone2022}.
In this work, we use such profiles as starting points for the reconstruction of the post-collision star (as described in Section~\ref{sec:postmerger_construction}). For comparison, we also compute two \parsec\ stellar tracks with an initial mass of  57.6~\Msun\ (hereafter, \textsc{p58}) and 88~\Msun\ (hereafter, \textsc{p88}), respectively. 
At 4.3 Myr, the \textsc{p58} star track has already ended the \ac{COB} phase and is just a few days from the final \ac{CC}.

\subsection{\textsc{mesa} models and pre-collision models}
\label{sec:mesa_conf}

For the \mesa\ evolutionary tracks, we adopted the same configuration as \citet[][hereafter R20]{renzo2020}\footnote{Publicly available at the following link \href{https://zenodo.org/record/4062493}{10.5281/zenodo.4062492}.}.  
All \mesa\ models are computed using a metallicity $Z=0.0002$.
Here, we describe the main differences in the adopted physics with respect to the \parsec\ configuration. We refer to R20 for a detailed description of the input physics of the \mesa\ models.
The \mesa\ stellar tracks are computed using the Ledoux criterion to define the convective borders \citep{Ledoux1947}, and the diffusion coefficient is calculated by adopting the mixing length theory framework with \amlt~=~2.0. We use the MLT++ artificial enhancement of the convective flux \citep{Paxton2015}. Overshooting is calculated by adopting the step scheme and following \citet{Brott2011}. Thermohaline mixing is included as in \citet{Farmer2016}. The pre-collision models are computed with an 8-isotope $\alpha$ chain nuclear network, while for post-collision models a 22-isotope network is adopted.

\begin{figure*}
    \includegraphics[width=\columnwidth]{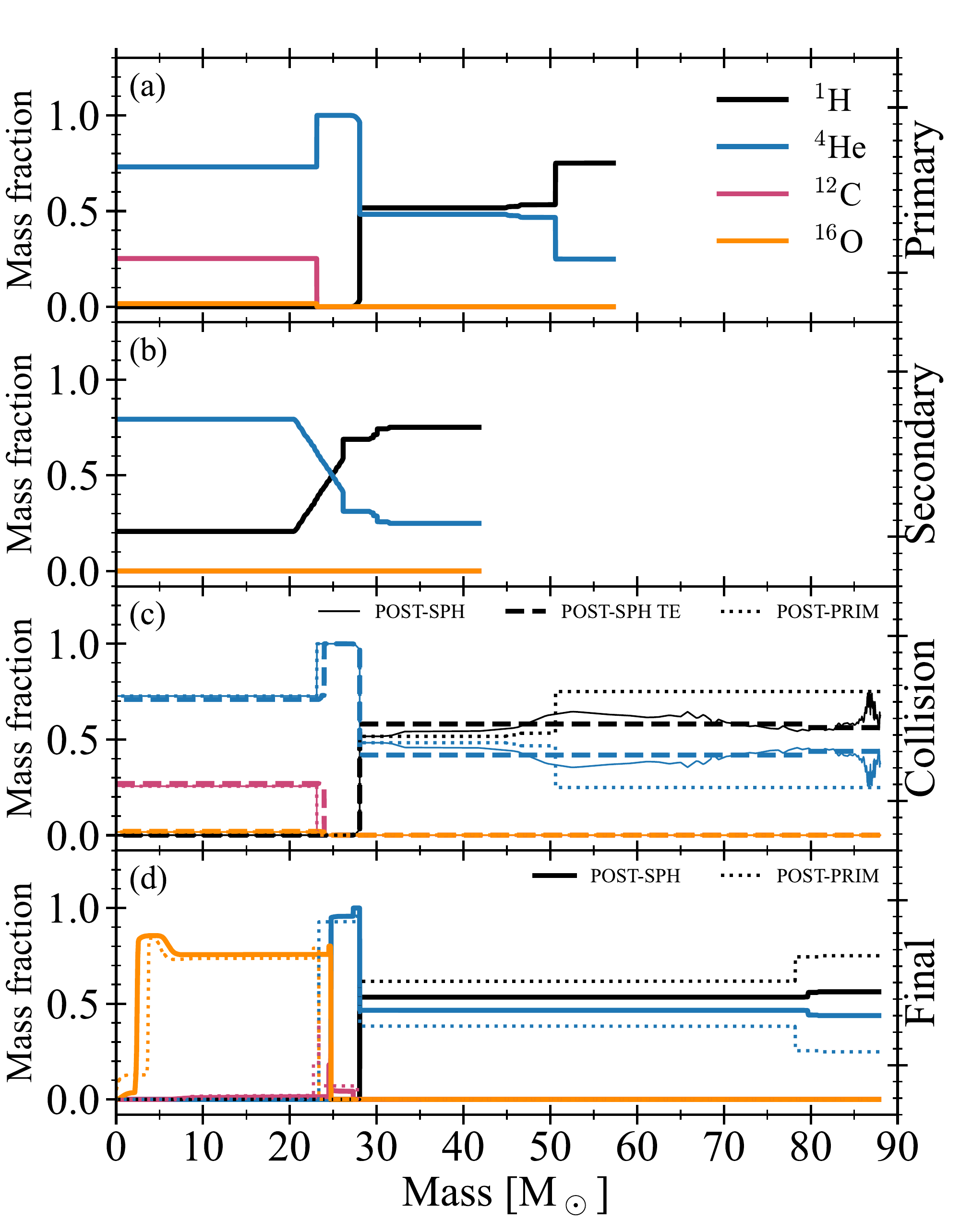}
    \includegraphics[width=\columnwidth]{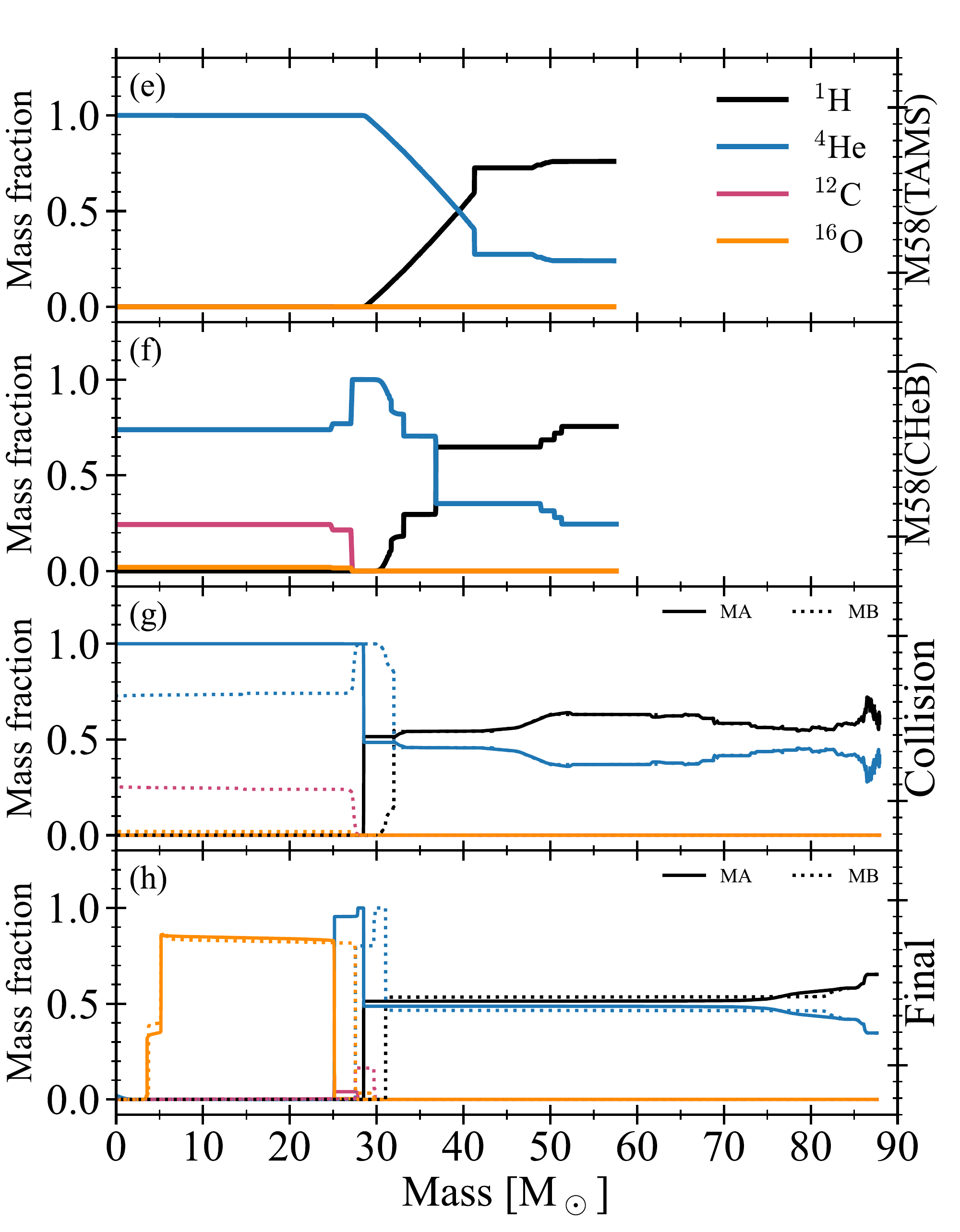}
    
    \caption{Chemical profiles in mass fraction of hydrogen (black), helium (blue), carbon (pink) and oxygen (orange) of different stellar models. Panels (a) and (b) show the \parsec\ primary and  secondary profiles before the collision, respectively. 
    In panel (c), the thin solid and dotted lines are the \textsc{post-sph} and \textsc{post-prim} models right after the collision, while the thick dashed line is the \textsc{post-sph} at thermal equilibrium (\textsc{post-sph te}). 
    Panel (d) shows the profiles of the post-collision \parsec\ tracks (\textsc{post-prim} and \textsc{post-sph}) at the end of \ac{COB}. Panels (e) and (f) show the profiles of the \mesa\ track \textsc{m58}, in the terminal age \ac{MS} and \ac{CHeB} phases, respectively. These two models are used to build the \mesa\ post-collision tracks \textsc{ma} and \textsc{mb}. 
    Panels (g) and (h) show the profiles of the post-collision \mesa\ tracks (\textsc{ma} and \textsc{mb}) just after the collision and at the onset of the \ac{CC}, respectively.
    }
\label{fig:profile}
\end{figure*}

We built two \mesa\ pre-collision models, taken from two different phases of the primary stellar track (hereafter, \textsc{m58}), that is computed from the \ac{ZAMS} to the end of \ac{CHeB}.
In the first case, we assume that the primary star collides while in the terminal age main sequence phase, as assumed by R20. The star is a \ac{BSG} with a radius of $\approx{45}$~\Rsun, and has a well-developed He-core of about 29~\Msun. Panel (e) of Fig.~\ref{fig:profile} shows the chemical profile of this model. 

In the second case, we assume that the primary star is in the \ac{CHeB} phase with a central helium $X_{\rm c}$(He) $\sim{0.74}$. At this stage, the primary is a \ac{RSG} with a radius of about 650~\Rsun\ and has a He-core of 30.4~\Msun. For computing the latter model, we suppressed the stellar winds since when the star approaches the \ac{RSG} stage they can remove several solar masses before we reach the desired stopping condition. In the \ac{RSG} phase, the mass loss could reach about 10$^{-4}$~\Msun yr$^{-1}$. By suppressing the winds, we can build the primary star in \ac{CHeB} with 57.6~\Msun. Panel (f) of Fig.~\ref{fig:profile} shows the chemical profile of this model.
We chose this evolutionary stage to make a comparison with the \parsec\ primary star model. 

\begin{table*} 
    \caption{List of stellar tracks.
    } 
    \begin{center}
    \begin{tabular}{lccc} 
        \hline\hline
        \multicolumn{4}{c}{\bf Single stellar tracks} \\
        \hline
        Model name & Code & \M{in}/\Msun & Description\\
        \hline
        \textsc{p58} & \parsec\ & 57.6 & Star evolved from \ac{ZAMS} to the end of \ac{COB} phase.\\
        \textsc{p88} & \parsec\ & 88 & Star evolved from \ac{ZAMS} to the end of \ac{COB} phase.\\
        \textsc{m58} & \mesa\ & 57.6 & Star evolved from \ac{ZAMS} to the end of \ac{CHeB} phase.\\
        \hline
        \multicolumn{4}{c}{\bf Post-collision stellar tracks} \\
        \hline
        \textsc{post-prim} & \parsec\ & 88 & The evolution starts as a \ac{RSG} in \ac{CHeB} phase with a primordial envelope composition.\\
        \textsc{post-sph} & \parsec\ & 88 &  The evolution starts as a \ac{RSG} in \ac{CHeB} phase, with the SPH simulation envelope composition.\\
        \textsc{ma}  & \mesa\ & 88 & The evolution starts as a \ac{BSG} star in the terminal age \ac{MS} phase. The same envelope composition as in the SPH simulation. \\
        \textsc{mb} & \mesa\ & 88 & The evolution starts as a \ac{BSG} star in the \ac{CHeB} phase. Same envelope composition as in the SPH simulation. \\
        \textsc{mix} & \mesa\ & 99 & Track with a complete mixed envelope from R20. \\
        \textsc{primordial} & \mesa\ & 99 & Track with a primordial envelope composition from R20. \\
        
        \hline
    \end{tabular}
   	\end{center}
    \flushleft{\footnotesize{
    The models \textsc{mix} and \textsc{primordial} are the mix and primordial models of R20, that we re-ran using their configuration.
        }}
\label{tab:models} 
\end{table*}

\subsection{Hydro-dynamical simulations}
\label{sec:hydro_sim}

As detailed in a companion paper \citep{Ballone2022}, we have run the hydrodynamical simulation of the collision between a \ac{CHeB} star and a \ac{MS} star, to infer the properties of their post-collision remnant. Such a simulation has been run with the \ac{SPH} code {\sc StarSmasher} \citep{gaburov2018}, which is particularly well-suited for simulations of stellar mergers. We generated the initial conditions for the colliding stars by importing the one-dimensional profiles obtained with {\sc parsec} (see Section \ref{sec:parsec}) and then re-mapped them into three-dimensional distributions of SPH particles.
To test the most extreme case in terms of mass lost in the collision, we put the stars on a radial orbit, with a velocity at infinity equal to 10 $\mathrm{km\; s^{-1}}$, corresponding to the typical velocity dispersion of young massive star clusters.

We evolved the stellar remnant with {\sc StarSmasher} until it reached the hydrodynamical equilibrium, and found that about 12\% of the total mass of the system is ejected during the collision, so that the post-coalescence star has a final mass equal to $87.9~\Msun$. We also found that in the encounter the MS star sinks down to the close vicinity of the He-burning core of the primary star, while the material lost mostly belongs to the outer H-rich envelope of the CHeB star. As a result, the chemical abundances re-arrange according to the profile shown in Fig.~\ref{fig:profile}. More details on the setup and results of this hydrodynamical simulation can be found in \citet{Ballone2022}. 
From this hydro-dynamical simulation we take the information about the mass loss and the final chemical profile\footnote{We do not use the full profiles derived from the hydro-dynamical simulation, because the resolution of the He core is not suitable for entropy relaxation with a stellar-evolution code \citep{Ballone2022}.}. We use this information to build a new stellar model by means of an accretion process, as detailed in the following section (see also R20).

\subsection{Constructing the post-collision model} 
\label{sec:postmerger_construction}

\subsubsection{\parsec}
To construct the post-collision star with \parsec,  
we accrete mass onto the 
primary star 
until its total mass 
becomes $\M{coll}=88$~\Msun. During accretion, we stop the chemical evolution and take into account the heat injected by the accreting material, as described by \cite{Kunitomo2017}, which leads the star to inflate and become a \ac{RSG}. Such expansion is expected from a head-on collision process \citep{sills1997, glebbeek2008, Glebbeek2013, Ballone2022}. More details on the accretion process are provided in the Appendix~\ref{a:accretion}. 
After mass accretion, we create two post-collision models. In the first model (hereafter \textsc{post-prim}) we maintain the pristine chemical composition of the envelope, while in the second model (\textsc{post-sph}) we change the chemical composition of the envelope to match the \ac{SPH} simulation by \citet{Ballone2022}. Panel (c) of Fig.~\ref{fig:profile} shows the chemical profiles of the two models, \textsc{post-prim} and \textsc{post-sph}. 
Both post-collision models have a helium core of $\sim 28$~\Msun, and an envelope of about 60~\Msun, by construction. The model \textsc{post-prim} has a total He content of 41.5~\Msun, while the model \textsc{post-sph} has 46.7~\Msun.
Then, we restart the chemical evolution and follow the evolution of the two stars until the end of the \ac{COB} phase.

\subsubsection{\mesa}

To construct the post-collision star with \mesa, we use the methodology described in R20. In the first step, the mass is accreted onto the primary model until the total mass reaches \M{coll}$=88$ M$_\odot$. Then, we changed the chemical profile of the envelope to reproduce the SPH simulation. During such building steps, the chemical evolution is stopped. In agreement with R20, but in contrast with the method we used for the \parsec\ models,  we do not include extra heating injection during the accretion. This methodology leads the models of the post-collision products to be luminous \ac{BSG}.

With the above methodology, we compute two \mesa\ post-collision tracks.
The first (hereafter \textsc{ma}) is built assuming that the primary star is in the terminal age \ac{MS} phase, therefore we start the accretion from the primary model \textsc{m58}(TAMS). The \textsc{ma} track right after the collision has a He-core of about 29~\Msun, and a total helium content of about 55~\Msun.

The second track (\textsc{mb}) is built assuming that the primary is in the \ac{CHeB} phase, then we start the accretion from the \textsc{m58}(CHeB) model. We accrete mass until the total mass reaches \M{coll}, and then change the chemical composition of the envelope. This post-collision model has a He-core of about 30~\Msun, a total helium content of about 47~\Msun, and a total carbon and oxygen content of about 7~\Msun. Panel~(g) of Fig.~\ref{fig:profile} shows the chemical profiles of the two reconstructed models with \mesa\ right after the collision.
We follow the evolution of the two stars until the onset of \ac{CC}.
The \emph{inlists} of the \mesa\ models and outputs are available at \href{https://zenodo.org/record/6418977}{10.5281/zenodo.6418976}. Table~\ref{tab:models} lists all computed tracks and their main properties.

\section{Results} 
\label{sec:Results}

\begin{figure*}
\includegraphics[width=\columnwidth]{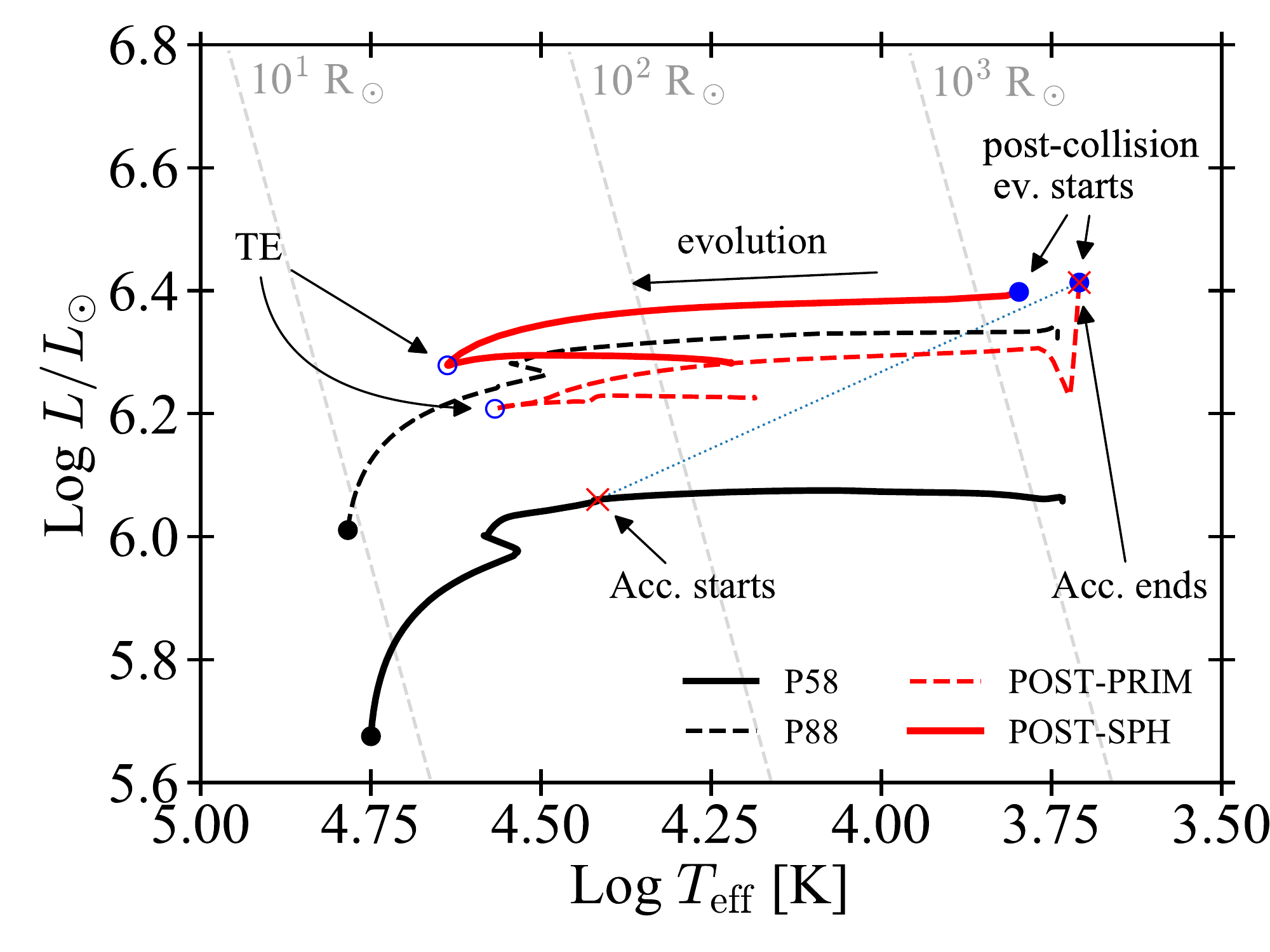} 
\includegraphics[width=\columnwidth]{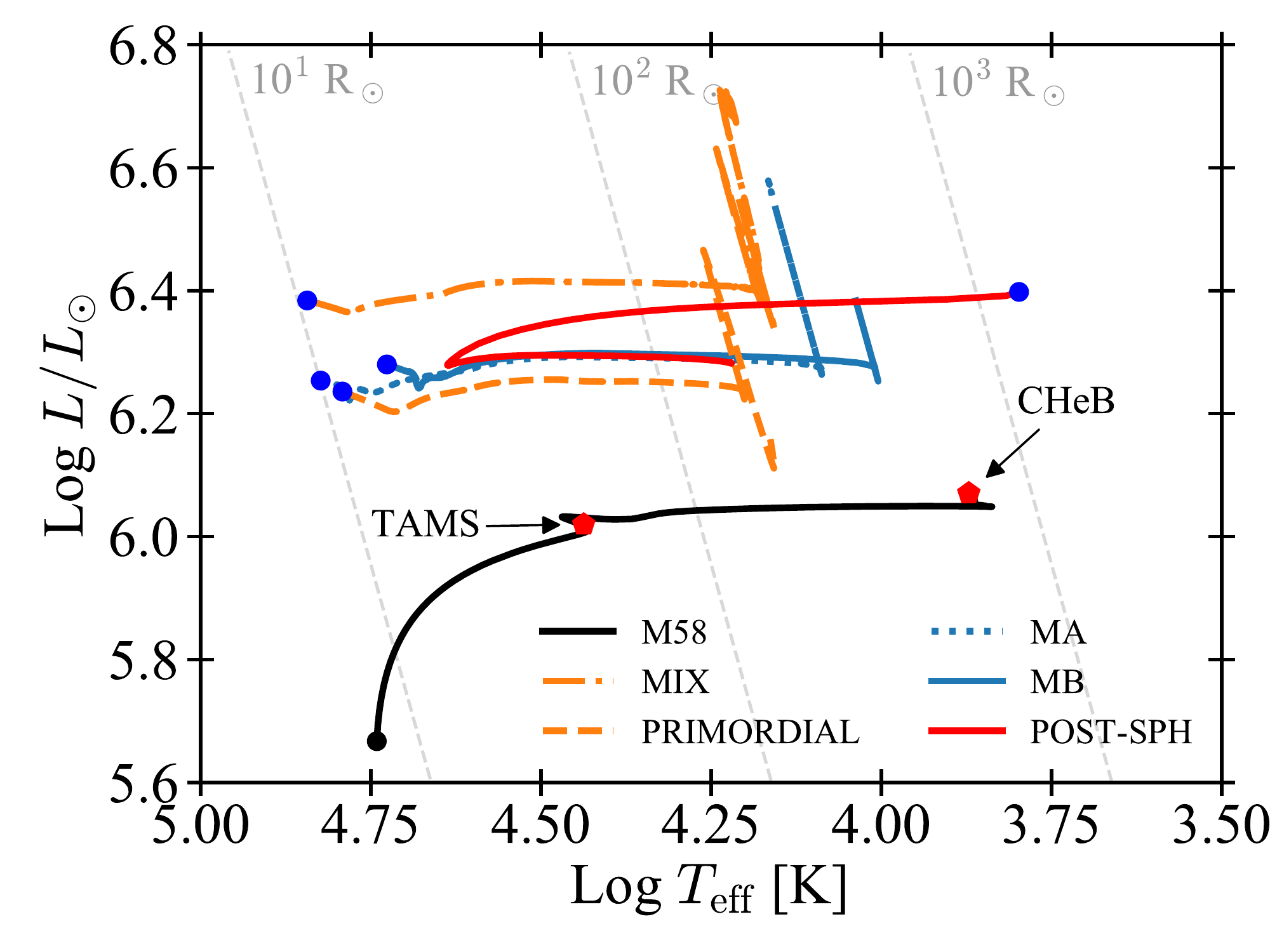}
    \caption{HR diagrams of the post-collision models.
    The left-hand panel shows the \parsec\ tracks. 
    The black points indicate the \ac{ZAMS} of the standard (single-star) tracks.
    The black solid line indicates the evolution of the primary star (\textsc{p58}). The red crosses mark the starting and ending points of the accretion process, which are connected by the thin blue dotted line. 
    The blue points indicate the starting point of the post-collision evolution tracks.
    The red dashed line shows the evolution of the post-collision track with pristine gas composition (\textsc{post-prim}).
    The red solid line shows the evolution of the \textsc{post-sph} track, in which we changed the chemical composition according to the \ac{SPH} simulation.
    The blue empty circles mark the point in which the post-collision tracks reach thermal equilibrium (TE).
    The black dashed line shows the evolution of the \textsc{p88} star track. 
    The right-hand panel shows the \mesa\ tracks. 
    The black point and the solid line indicate the \ac{ZAMS} and the evolution of the primary track \textsc{m58}.
    The blue points indicate the starting point of the post-collision evolution tracks.
    The  dashed and dot-dashed orange lines are the \textsc{mix} and \textsc{primordial} models from R20, respectively (the \textsc{mix} model assumes complete chemical mixing, while the \textsc{primordial} model assumes no mixing).
    The dotted and solid blue lines are the models \textsc{ma} and \textsc{mb}, respectively. 
    The solid red line is the \parsec\ track \textsc{post-sph}, for comparison.
    The two red pentagons mark the two evolutionary stages we use to construct the \mesa\ post-collision models \textsc{ma} and \textsc{mb}, terminal age \ac{MS} (TAMS) and \ac{CHeB}. 
    }
\label{fig:HR}
\end{figure*}
\begin{figure*}
\includegraphics[width=\columnwidth]{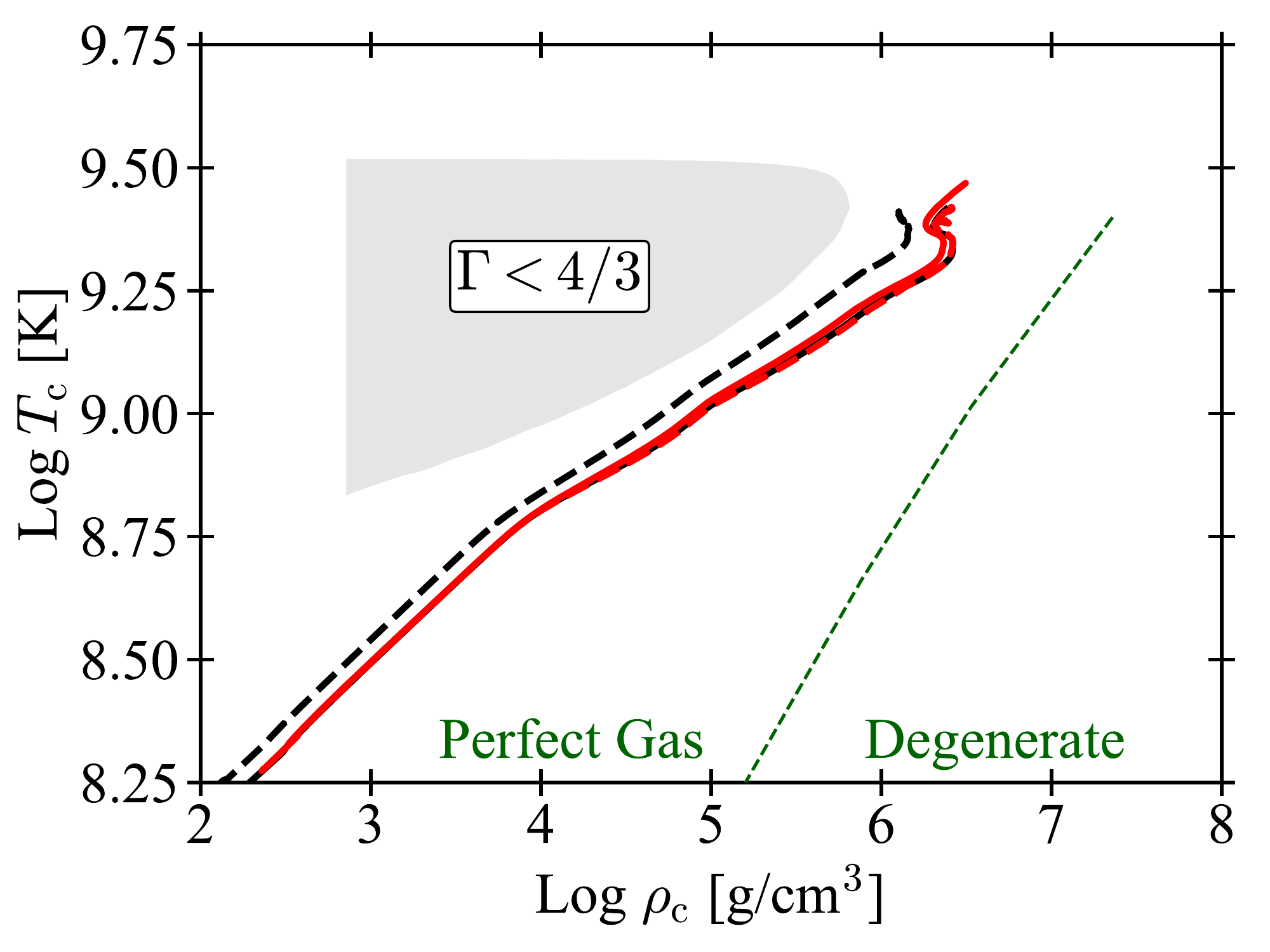}
\includegraphics[width=\columnwidth]{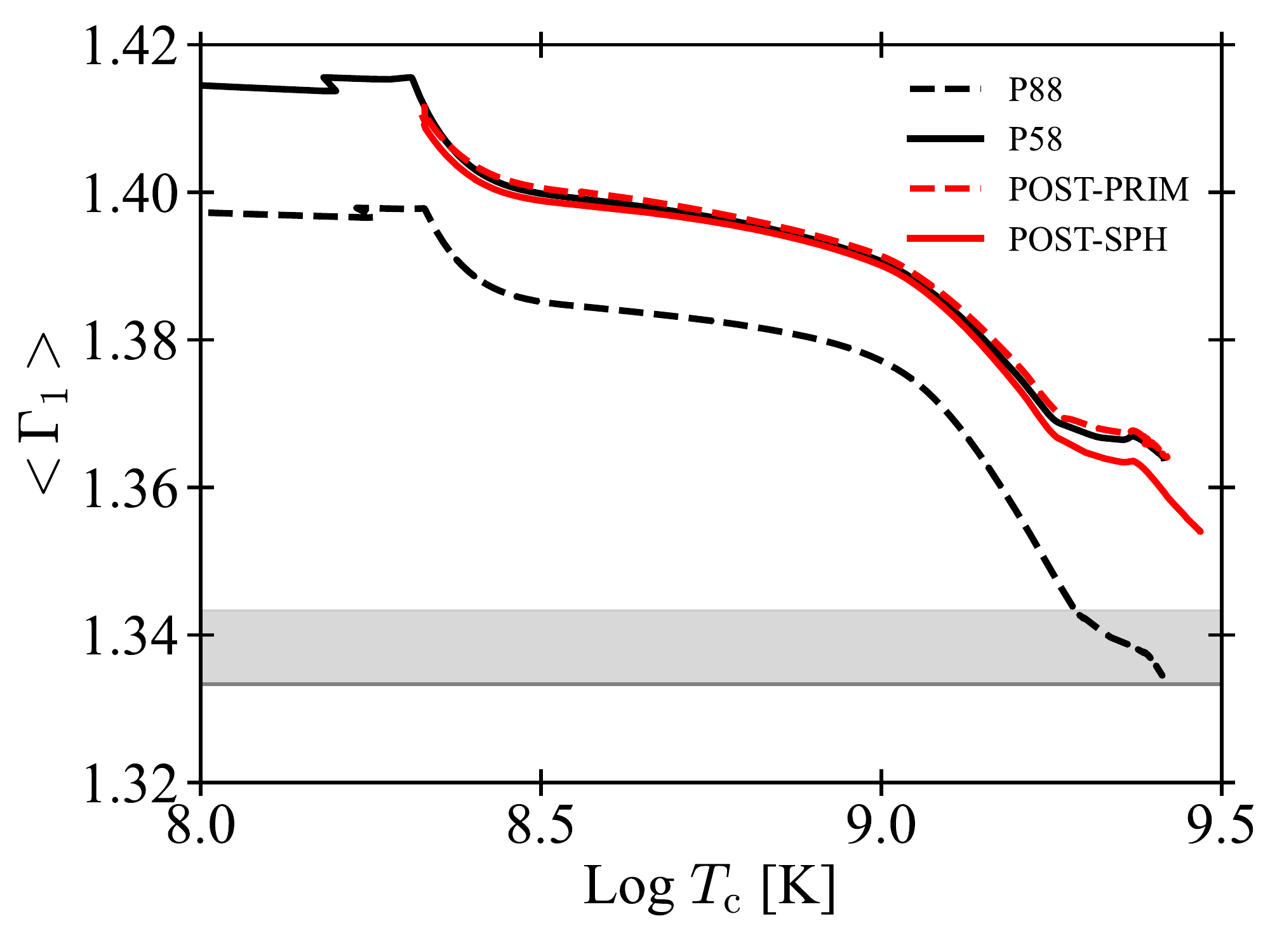}
    \caption{The left-hand panel shows the evolution of the central temperature versus central density of the post-collision tracks.  
    The right-hand panel shows the evolution of \Gav\ as a function of the central temperature. 
    In both panels, black dashed line: track {\sc p88}; black solid line: {\sc p58}; red dashed line: {\sc post-prim}; red solid line: {\sc post-sph}.
    The grey area in the left-hand panel indicates the instability region with $\Gamma_1 < 4/3$. 
    In the right-hand panel, the thin dark gray horizontal line corresponds to  $\langle{}\Gamma_1\rangle{} = 4/3$. The above light-gray area indicates values of  \Gav\ between 4/3 and 4/3 + 0.01, i.e. the range in which the whole star becomes dynamically unstable \citep{farmer2019, costa2021}. The post-collision tracks (\textsc{post-sph} and \textsc{post-prim})  almost overlap with the track \textsc{p58} (with mass 57.6 M$_\odot$) in both diagrams. 
    }
\label{fig:rhocTc_PI}
\end{figure*}
\begin{figure*}
\includegraphics[width=\columnwidth]{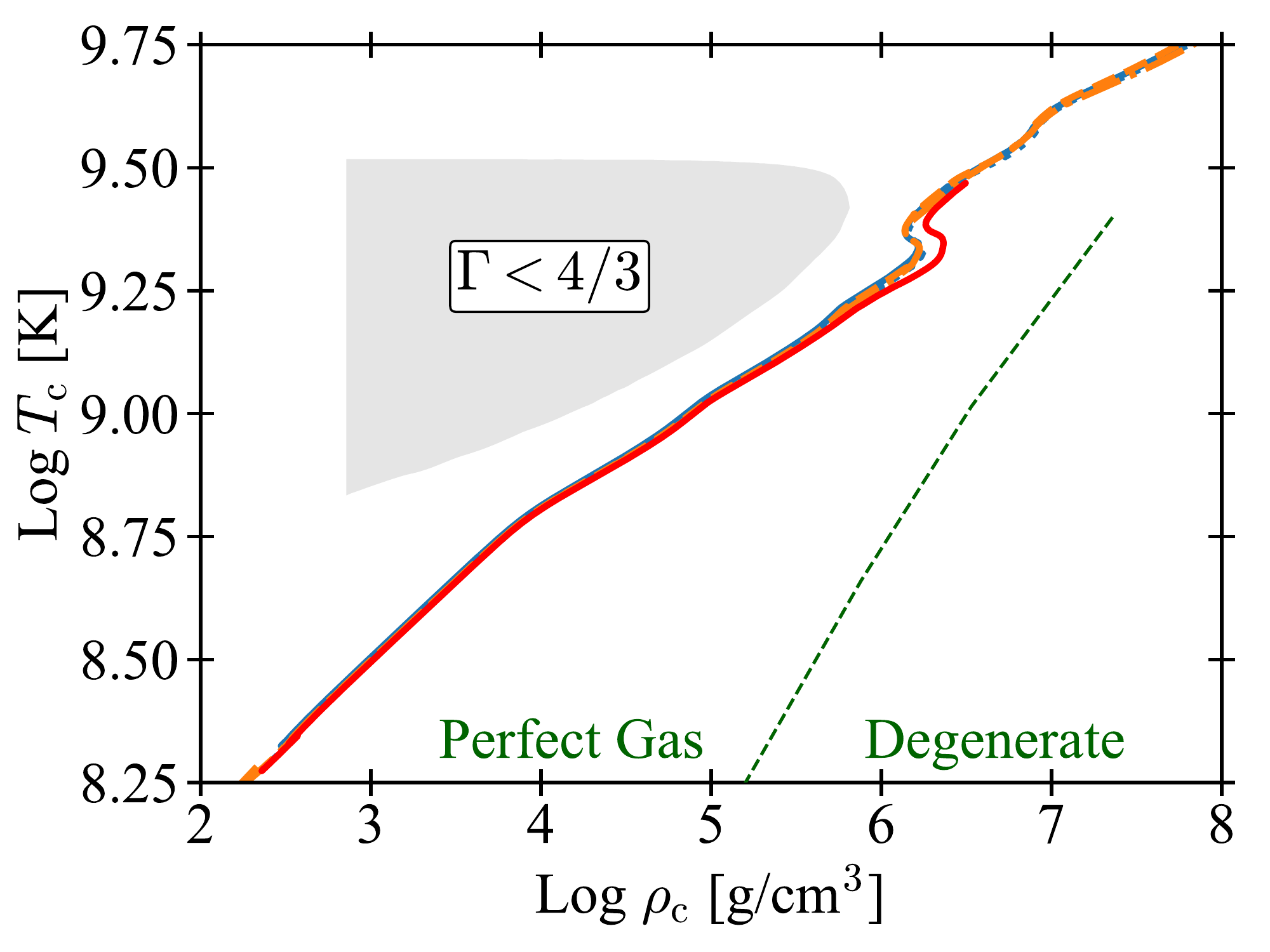}
\includegraphics[width=\columnwidth]{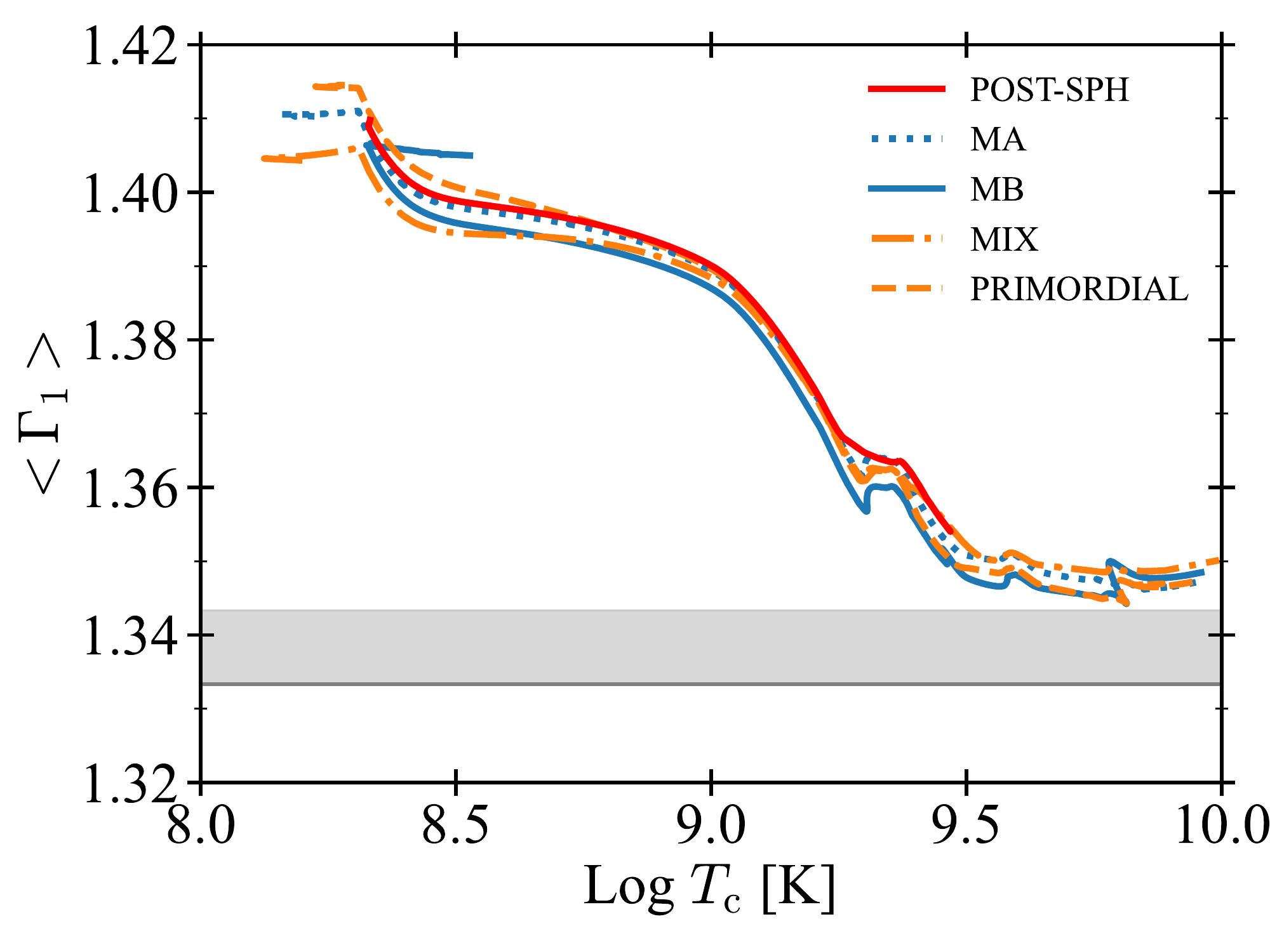}
    \caption{Comparison between \parsec\ and \mesa\ tracks. Left-hand panel: 
    central temperature versus central density; 
    right-hand panel:  \Gav\ versus the central temperature. In both panels, the red solid line is the \parsec\ {\sc post-sph} track, while the blue dotted and dashed lines are the \mesa\ tracks \textsc{ma} and \textsc{mb}, respectively. The orange  dot-dashed and dashed lines are the R20 post-collision tracks \textsc{mix} and \textsc{primordial}, respectively. 
    }
\label{fig:rhocTc_PI_mesa}
\end{figure*}
\begin{figure}
\includegraphics[width=\columnwidth]{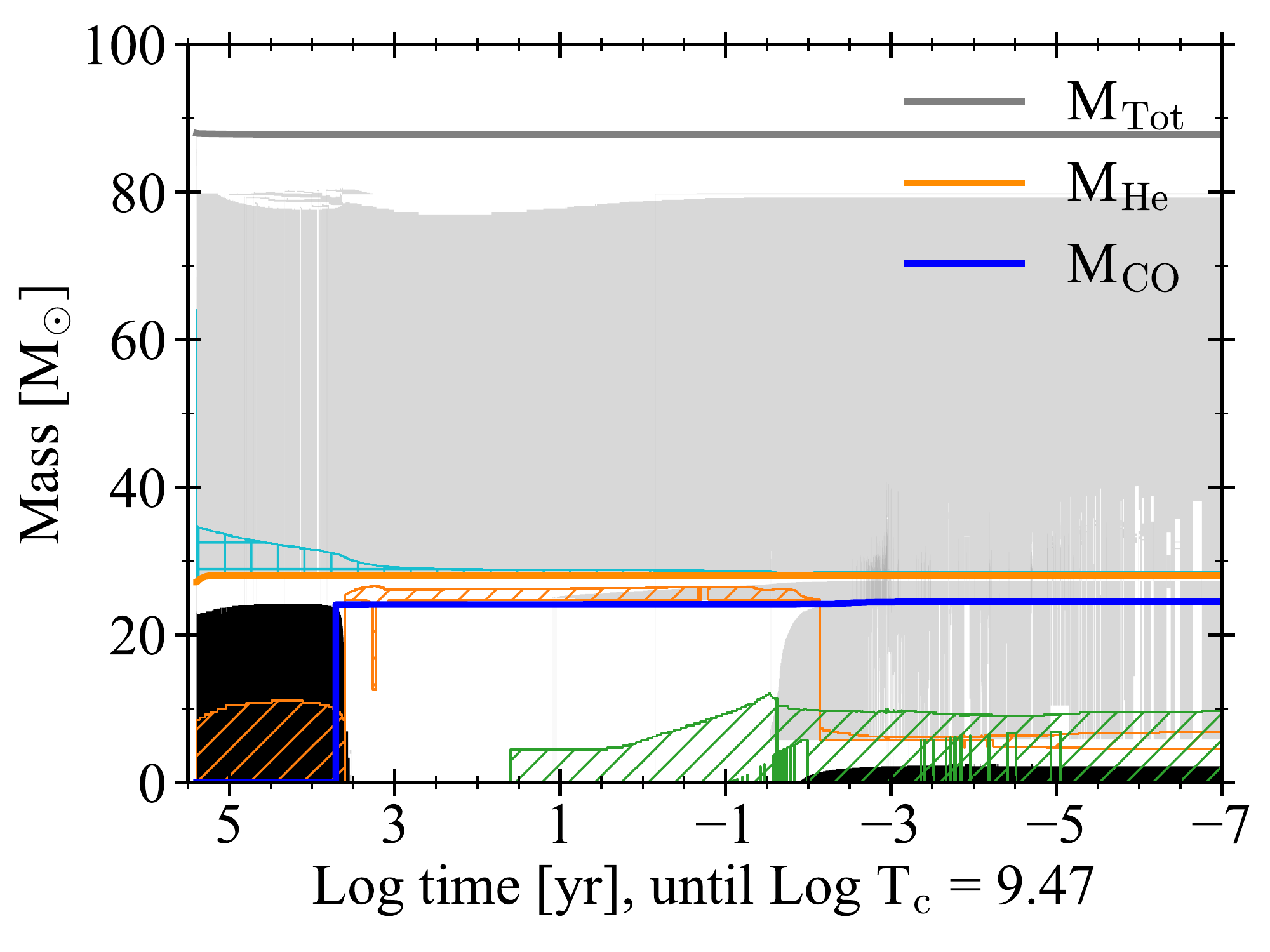}
    \caption{Kippenhahn diagram of the post-collision stellar track \textsc{post-sph}. The black area is the convective core, the grey areas indicate the convective envelope and intermediate convective regions. Continuous gray, yellow, and blue lines show the total stellar mass, He core mass and  CO core mass, respectively. The cyan, yellow, and green hatched areas indicate the hydrogen, helium and carbon burning regions, respectively. We plot zones that contribute at least for the 1\% to the H, He and C luminosity at a given time-step.
    }
\label{fig:Kipp}
\end{figure}

\subsection{Post-collision evolution}
\label{sec:post-coll-ev}
Figure~\ref{fig:HR} shows the evolution of the post-collision stellar tracks in the \ac{HR} diagram. The left-hand panel shows the comparison of \textsc{post-prim} and \textsc{post-sph} with the "standard" stellar tracks \textsc{p58} and \textsc{p88}. 
Both post-collision tracks start their evolution as \ac{RSG}s. During the \ac{RSG} phase, they undergo a relatively high mass loss ($2 \times 10^{-5}$~\Msun yr$^{-1}$) by stellar winds. They rapidly evolve toward the blue part of the \ac{HR} diagram, reaching thermal equilibrium in $\sim 8\times10^3$~ yr and $\sim 5.5\times10^3$~ yr for \textsc{post-prim} and \textsc{post-sph}, respectively. This time scale is several times the Kelvin-Helmholtz time-scale of the models at the beginning of their evolution, that is, about 100~yr. When they reach thermal equilibrium, they have lost less than 0.05~\Msun, and restart burning helium in a stable way.
Panel (c) of Fig.~\ref{fig:profile} shows the profile of the track \textsc{post-sph} in TE. During thermal relaxation, the star develops large convective regions which rapidly homogenize the envelope. The small inversion of molecular weight in the post-collision model is canceled by convection. Therefore, mixing processes due to molecular weight inversions (i.e., thermohaline mixing) do not take place during the evolution. 
In this phase, both models still have a central helium fraction of about $X_{\rm c}\mathrm({\rm He}) \sim 0.7$.
The higher luminosity of the \textsc{post-sph} track compared to the \textsc{post-prim} track depends mainly on the different helium content in the stellar envelope and surface. Generally, stars with a stellar surface enriched with He are more luminous, due to the higher mean molecular weight \citep{Kippenhahn2012}.

After thermal relaxation, the stellar tracks \textsc{post-sph} and \textsc{post-prim} are \ac{BSG}s, with a radius of about 25~\Rsun\ and 31~\Rsun, respectively. The mass loss by stellar winds is quenched ($\sim 3 \times 10^{-7}$~\Msun~yr$^{-1}$).
During the \ac{CHeB} phase, the stars move to the red part of the \ac{HR} diagram. 
The \ac{CHeB} phase of the two stars lasts for 
$\sim 2.4\times10^5$~ years for \textsc{post-prim} and $\sim 2.2\times10^5$~years for \textsc{post-sph}. 
After \ac{CHeB}, both stars evolve until the end of \ac{COB} in $ \sim 4 \times10^3$ years and end their life as \ac{BSG}s.
These stars retain most of their envelope during evolution due to 
inefficient line-driven winds \citep{vink2021}.
In contrast, the standard tracks \textsc{p58} and \textsc{p88} end their life as \ac{RSG}s. 
The different final stellar configuration may have an impact on the final mass of the \ac{BH}, since \ac{RSG} stars may shed their envelope during core collapse, because of neutrino loss \citep[][]{Fernandez2018}. 

The right-hand panel of Fig.~\ref{fig:HR} compares the post-collision \mesa\ tracks and the \textsc{post-sph} track. In the same Figure, we added the \textsc{mix} and \textsc{primordial} post-collision tracks from R20 for comparison. 
By construction, the \mesa\ tracks begin their evolution as \ac{BSG}s. After a brief expansion phase, they start the \ac{CHeB} phase and move to lower effective temperatures. All \mesa\ models end their life as \ac{BSG}s.
This Figure shows that the post-collision stellar tracks with the total mass equal to \M{coll} and the reconstructed stellar envelope from the \ac{SPH} simulation have a very similar evolutionary path, even if they have been built with a different methodology and simulated with a different stellar evolution code. The \mesa\ tracks tend to end their life with larger stellar radii than the \parsec\ tracks. Table~\ref{tab:results} shows the properties of the stellar tracks at the end of \ac{COB} and at the onset of \ac{CC} for the \mesa\ models.

Figure~\ref{fig:rhocTc_PI} shows the evolution of the central temperature (\Tc), density (\rhoc), and pressure-weighted average adiabatic index \citep[\Gav,][]{farmer2020, Renzo2020a, costa2021} of the \parsec\ tracks. 
The tracks \textsc{post-prim} and \textsc{post-sph} almost overlap with track \textsc{p58} in the figure, because they have a very similar He core.
These models avoid \ac{PI} (i.e., $\langle{}\Gamma_1\rangle{} > 4/3$ during the whole evolution) and will eventually undergo \ac{CC}. 
The final \ac{BH} mass depends on the pre-supernova total mass and on the fate of the envelope during collapse (as discussed in Sect.~\ref{sec:Discussion}).

The 'standard' \parsec\ stellar track \textsc{p88}, builds up a bigger He core during the \ac{CHeB} phase ($M_{\rm He}\approx{46}$~\Msun) with respect to the exotic post-collision stars, and becomes unstable to PI during the \ac{COB} phase \citep[in an off-center region,][]{woosley2019, farmer2020, Renzo2020a, costa2021}. 

Figure~\ref{fig:rhocTc_PI_mesa} shows the evolution of \Tc, \rhoc, and \Gav\ for the \mesa\ tracks. The evolutionary paths of these stellar tracks are very similar, as they all evolve from the \ac{CHeB} phase to the final \ac{CC} avoiding \ac{PI}. 
This Figure shows that the \textsc{post-sph} track evolves in a similar way to the \mesa\ tracks (\textsc{ma} and \textsc{mb}). 

\begin{table*} 
    \caption{Properties of the models at the end of \ac{COB}, and at the onset of \ac{CC}.} 
    \begin{center}
    \begin{tabular}{lccccccccccc} 
        \hline\hline
        \multicolumn{10}{c}{End of core oxygen burning} \\
        \hline
        Model & $M_*$/M$_\odot$ & $R_*$/R$_\odot$ & Log $L/$L$_\odot$ &  Log $T_\mathrm{eff}$/K &  Log $g$/(cm s$^{-2}$)  & \M{He}/M$_\odot$ & \M{CO}/M$_\odot$ & $\xi_{2.5}$   &  $\xi_\mathrm{env}$ & $M_\mathrm{ej}$/M$_\odot$ & $M_\mathrm{BH}$/M$_\odot$ \\
        \hline
        \textsc{post-prim}  & 87.9 & 183.7 & 6.227 & 4.187 & 1.854 & 28.1 & 23.3 & 0.250 & 0.479  &  0.04 - 0.28 &  87.3\\
        \textsc{post-sph}   & 87.8 & 167.2 & 6.284 & 4.221 & 1.935 & 28.1 & 24.5 & 0.279 & 0.525  &  0.03 - 0.26 &  87.3\\
        \textsc{mix}        & 98.5 & 268.1 & 6.687 & 4.219 & 1.575 & 29.2 & 25.4 & 0.293 & 0.367  &  -    &  -  \\
        \textsc{primordial} & 98.9 & 187.1 & 6.138 & 4.160 & 1.889 & 28.9 & 25.8 & 0.273 & 0.528  &  -    &  -  \\
        \textsc{ma}         & 87.6 & 301.5 & 6.529 & 4.154 & 1.422 & 28.5 & 25.1 & 0.277 & 0.291  &  -    &  -  \\
        \textsc{mb}         & 87.7 & 436.7 & 6.258 & 4.006 & 1.100 & 31.0 & 27.5 & 0.281 & 0.201  &  -    &  -  \\
        \hline\hline
        \multicolumn{10}{c}{Onset of core collapse} \\
        \hline
        Model & $M_*$/M$_\odot$ & $R_*$/R$_\odot$ & Log $L/$L$_\odot$ &  Log $T_\mathrm{eff}$/K &  Log $g$/(cm s$^{-2}$)  & \M{He}/M$_\odot$ & \M{CO}/M$_\odot$ &  $\xi_{2.5}$  &  $\xi_\mathrm{env}$ & $M_\mathrm{ej}$/M$_\odot$ & $M_\mathrm{BH}$/M$_\odot$\\
        \hline
        \textsc{mix}        & 98.5 & 271.0 & 6.701 & 4.220 & 1.562 & 29.2 & 25.6 & 0.510 & 0.362  &  0.03 - 0.27 &  97\\
        \textsc{primordial} & 98.9 & 188.9 & 6.158 & 4.163 & 1.880 & 28.9 & 25.9 & 0.561 & 0.523  &  0.03 - 0.22 &  97.3\\
        \textsc{ma}         & 87.6 & 302.9 & 6.575 & 4.165 & 1.418 & 28.5 & 25.1 & 0.591 & 0.289  &  0.04 - 0.36 &  87\\
        \textsc{mb}         & 87.7 & 436.7 & 6.384 & 4.037 & 1.101 & 31.0 & 27.5 & 0.541 & 0.201  &  0.06 - 0.43 &  87\\
        \hline
    \end{tabular}
   	\flushleft{
    \footnotesize{From left to right: track name, stellar mass, radius, luminosity, effective temperature, surface gravity, He core mass, CO core mass,} core compactness, envelope compactness,  estimated ejected mass in a failed supernova, and expected \ac{BH} mass. The upper part of the Table shows the values of these quantities at the end of \ac{COB}, while the lower part indicates the values at the onset of \ac{CC}. To compute the final mass of the R20 models (\textsc{mix} and \textsc{primordial}) we take into account the luminous blue variable like outbursts assuming $M_\mathrm{LBV} = 1$~\Msun.
        }
        \end{center}
\label{tab:results} 
\end{table*}

Figure~\ref{fig:Kipp} shows the evolution of the internal structure of the track \textsc{post-sph}. 
During the \ac{CHeB} phase, the star has a convective core that extends up to about 28~\Msun, and a large intermediate convective region in the envelope fueled by the hydrogen-burning shell. After \ac{CHeB}, core convection is quenched, and the core contracts while the envelope expands. In this phase, the star has a double burning shell in which helium burns above the CO core, while hydrogen burns above the He core. 
About 70~yr before the final collapse, the star ignites carbon in the core. When central carbon is depleted, the carbon burning shell moves outward and triggers an internal convective zone. 
The energy provided by the carbon-burning shell sustains the external layers of the star and prevents \ac{PI} \citep{woosley2019, farmer2020, costa2021}. 
In this phase, neon photo-disintegration occurs, and when the central temperature reaches $\sim 2$~GK, oxygen is ignited ($\sim 6$ days before the final collapse). Oxygen burns in a small convective core of about 2.2~\Msun. At the end of the computation (that is, when the central oxygen is depleted), the central core is composed of 2.8\% of $^{23}$Na, 38\% of $^{28}$Si, 36\% of $^{32}$S, 10\% of $^{36}$ Ar, and 13\% of $^{40}$Ca. After this stage, in a few hours, the star will burn silicon and undergo the final \ac{CC}.

\subsection{Core and envelope compactness} 
\label{sec:compactness}

We compute the core compactness, $\xi_{2.5}$, of the last model of our simulations following \citet{OConnor2011} and  references therein: 
\begin{equation}
    \xi_{2.5} = \frac{2.5}{r(M = 2.5~\Msun)/10^3 \mathrm{km}},
    \label{eq:core_compact}
\end{equation}
where $r(M = 2.5~\Msun)$ is the radius that encloses 2.5~\Msun. 
We also computed the envelope compactness, $\xi_{\mathrm{env}}$, which is a proxy for the surface gravity of the star:
 \begin{equation}
    \xi_\mathrm{env} = \frac{M_*}{R_*},
    \label{eq:env_compact}
\end{equation}
where $M_*$ and $R_*$ are the total mass and radius of the star. We compute the compactness of the core and envelope at the end of the \ac{COB} phase (which corresponds to the end of the simulation for \parsec\ models), and at the onset of \ac{CC}. The results are shown in the upper and lower panels of Table~\ref{tab:results}, respectively.

\citet{Fernandez2018} have shown that mass ejection can occur even in a failed supernova, because of the instantaneous loss of neutrinos triggering a shock. Extending the previous work of \citet{Lovegrove2013} and \citet{Lovegrove2017}, \cite{Fernandez2018} found that mass ejection can occur in all types of \ac{BH} progenitors. 
The energy ejected from the explosion is a decreasing function of $\xi_{2.5}$, while the ejected mass is a decreasing function of $\xi_{\mathrm{env}}$.

\begin{figure}
\includegraphics[width=\columnwidth]{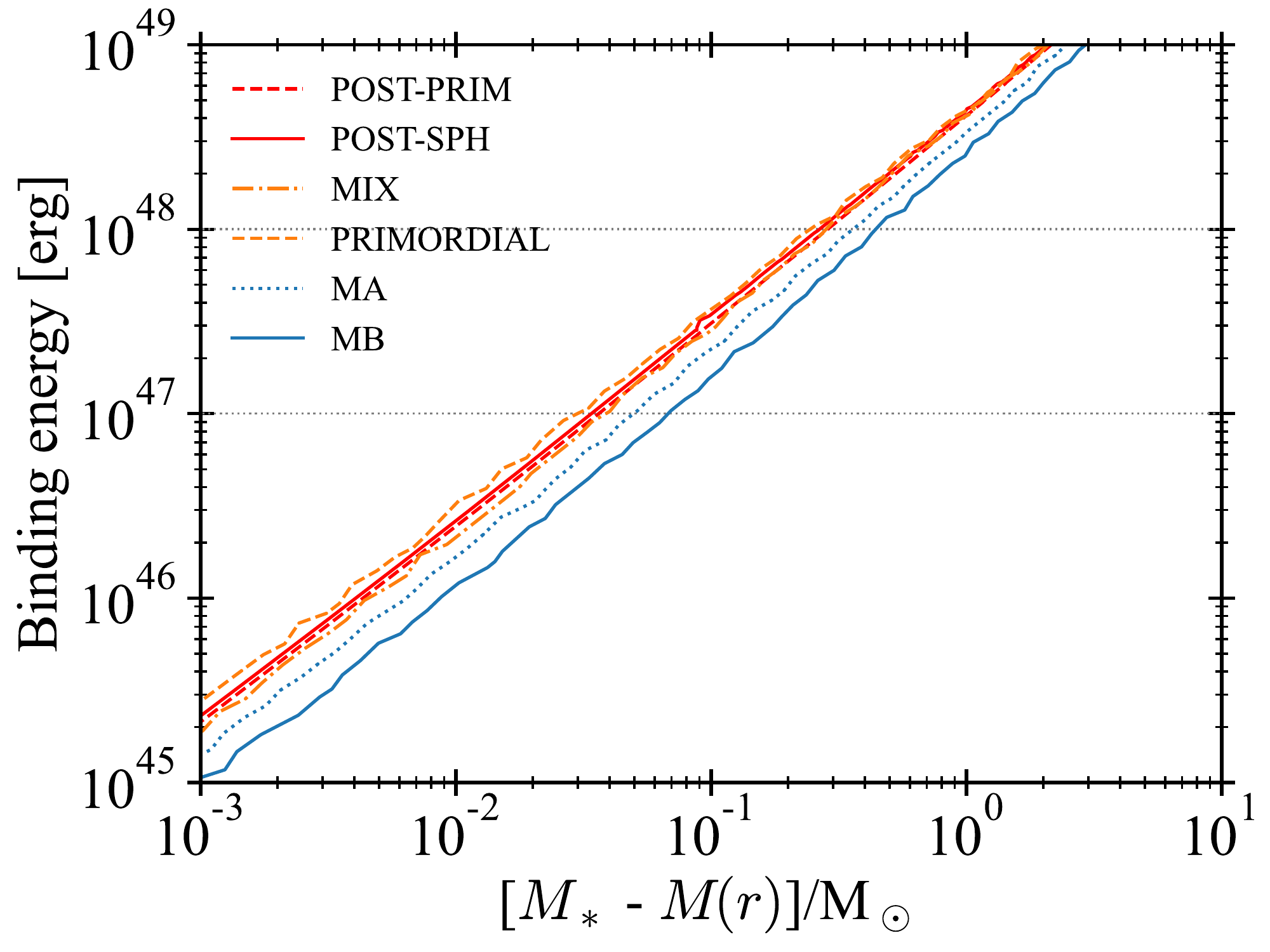}
\caption{Binding energy of the external part of the stellar envelope, from the exterior to the inner structure of our models, at the end of the evolution. The horizontal dotted gray lines show the minimum and maximum values of energy ejected (E$_{\rm ej}$) in a failed supernova for \ac{BSG} stars, as found by \protect{\citet{Fernandez2018}} 
}
\label{fig:Bind_energ}
\end{figure}

Figure~\ref{fig:Bind_energ} shows the binding energy of the external parts of the envelope at the end of the \ac{COB} phase for \parsec\ tracks, and at the onset of the \ac{CC} for \mesa\ tracks. All our models die as \ac{BSG} stars, with envelope compactness between $\xi_{\rm env}=0.2$ and 0.5 (as shown in Table~\ref{tab:results}). 
To give an estimate of the mass ejected by neutrino loss, we assume two extreme cases of ejected energy during the \ac{CC}, from the results of \citet{Fernandez2018}. Assuming an instantaneous loss of gravitational mass $\delta M_\mathrm{G} \sim 0.3$~\Msun, \cite{Fernandez2018} found an ejected energy between $10^{47}$ and $10^{48}$ erg for \ac{BSG}s. The corresponding ejected mass ($M_\mathrm{ej}$) in the two cases goes from 0.03~\Msun\ to 0.43~\Msun\ for the most and least compact of our models, respectively (Table~\ref{tab:results}).

In the final mass estimations, we did not include possible mass loss during \ac{CC} due to a shock revival that propagates outward after  core bounce \citep[][and references therein]{Powell2021, Rahman2022}. In the case of pulsational-pair instability progenitors of \ac{BH}s, \citet{Rahman2022} found that the shock (or sonic pulse) propagating up to the stellar surface may lead to an upper limit of the unbound mass that goes from 0.07~\Msun\ to 3.5~\Msun, depending on the progenitor structure. We cannot straightforwardly apply such mass loss  to our models, because of the different progenitor physical configuration. In the case of \cite{Rahman2022}, the \ac{BH} progenitor is a stripped star which suffered several episodes strong of mass loss due to the pulsations induced by pair creation.

\section{Discussion: the final BH mass} 
\label{sec:Discussion}

We integrated the evolution of an exotic star with a small core and a massive envelope) formed by stellar collision. 
Such type of \ac{BH} progenitors may also form via single stellar evolution depending on the stellar physics adopted. The boundaries of the \ac{PI} mass gap are very sensitive to stellar convection treatment, nuclear reaction rates, and stellar winds  \citep[][]{farmer2020, costa2021, vink2021}. Here, we investigate in detail the collision scenario, using a standard configuration for the stellar evolution.

Our stellar models come from the results of a hydrodynamical simulation \citep{Ballone2022}, which shows that the mass lost during the collision is not negligible, corresponding approximately to 12\% of the total mass of the system. 
All our models avoid \ac{PI} and evolve until the final \ac{CC} (Fig.~\ref{fig:rhocTc_PI_mesa}). This happens because the collision leaves the core of the primary star nearly untouched. 

We have calculated the core and envelope compactness ($\xi_{2.5}$ and $\xi_{\rm env}$) at the end of the simulations, to have a rough estimate of the mass lost during the final collapse, because of shocks induced by neutrino emission \citep{Fernandez2018}. 
We find that $\xi_{\rm env}$ does not change after the \ac{COB} phase. This happens because after \ac{COB} the evolution of the core is very fast and almost detached from the evolution of the envelope. Therefore, we can estimate the mass ejected during the failed supernova ($M_\mathrm{ej}$) also from models that do not evolve until the onset of \ac{CC}. 
In contrast, the core compactness changes significantly during the last phases. Thus, we cannot estimate the energy ejected during collapse from the models computed up to the end of \ac{COB} \citep{chieffi2021}. 

All post-collision stellar tracks cross the S Doradus instability strip during the \ac{CHeB} or core carbon burning phases and evolve beyond the Humphrey-Davidson limit. 
In such regions, a star may suffer from opacity-driven outburst episodes, typical of luminous blue variable stars. 
R20 found that stellar merger products may live for hundreds of years in proximity to the Eddington luminosity limit, and estimated that the total mass lost from the outbursts ($M_\mathrm{LBV}$) is about 1~\Msun, for both the \textsc{mix} and \textsc{primordial} tracks. 
Our \mesa\ models reach the proximity of the Eddington limit just at the end of their evolution ($\sim 1$~day before \ac{CC}); hence, we expect a negligible mass loss from such an opacity-driven process.

Based on these considerations, we now give an estimate of the final \ac{BH} mass ($M_\mathrm{BH}$) of the post-collision tracks. Starting from the pre-supernova mass of the star ($M_*$), we subtract the mass lost because of shocks induced by neutrino emission in a failed supernova \citep{Fernandez2018} as follows:
\begin{equation}
    M_\mathrm{BH} = M_* - \delta M_\mathrm{G} - M_\mathrm{ej},
    \label{eq:BH_mass}
\end{equation}
where $\delta{M}_{\rm G}=0.3$ M$_\odot$ is the instantaneous loss of gravitational mass and $M_{\rm ej}$ is the ejected mass because of neutrino-driven shocks.
The last column of Table~\ref{tab:results} lists the estimated final \ac{BH} mass for each track. For this computation, we use the maximum case of \M{ej}. For example, in the case of the track \textsc{post-sph}, we use \M{ej} = 0.26 M$_\odot$.
This result implies that the final mass of the BH  will be about 87~\Msun\ (\textsc{post-sph} model), lying inside the PI mass gap. 

In our study, we did not include the effect of possible chemical asymmetries due to the collision of the two stars, as found by \citet{Ballone2022}. Such asymmetries are not followed in our 1d models, but  we  expect that they do not affect the post-collision evolution.
In fact, the \textsc{parsec} post-collision tracks, which take into account the energy injected during accretion, become \ac{RSG}s and develop large convective envelopes. Such convective regions would mix eventual chemical asymmetries in a timescale much faster than the evolutionary timescale of the star (as described in Sect.~\ref{sec:post-coll-ev}). This suggests that eventual chemical asymmetries should be rapidly homogenized in the stellar envelope. 
In \citet{Ballone2022} the two stars collide head-on, forming a non-rotating post-collision stellar product.
In the case of an off-center collision, the post-collision product will be a rotating star. Rotation may lead to a geometrical distortion, to a growth of the stellar core, and to an enhancement of the mass loss. Depending on the amount of angular momentum in the star after the collision and on its evolutionary stage, rotation may change the star's final fate. We will investigate the role of rotation in forthcoming papers.

\section{Summary} 
\label{subsec:Conclusions}

We have studied the evolution of
a post-collision star, by means of detailed stellar evolution calculations with the \parsec\ and \mesa\ codes.
To reconstruct the post-collision star, we used results from the \ac{SPH} simulation by \citet{Ballone2022},  taking into account the mass lost during the collision (12\% of the total mass) and the new chemical profile. 

We find that the stellar tracks computed for the post-collision stars 
avoid \ac{PI} and evolve until the final \ac{CC}. Remarkably, the \parsec\ and \mesa\ stellar models evolve in a very similar way, ending their life as \ac{BSG} stars. 
We estimate the final \ac{BH} mass by taking into account the possible mass ejected during the final collapse, due to shocks induced by neutrino loss; 
we find that all of our models lose less than 0.5~\Msun\ during the final collapse, because of the relatively high compactness of the stellar envelope ($\xi_{\rm env}=0.4-0.5$). 

Thus, we expect that all our models produce \ac{BH}s with mass $\approx{87}$~M$_\odot$, within the \ac{PI} mass gap.

\section*{Acknowledgements}

We thank Morgan MacLeod and Mathieu Renzo for useful discussions. A. Ballone, G. Costa and M. Mapelli acknowledge financial support from the European Research Council for the ERC Consolidator grant DEMOBLACK, under contract no. 770017. 
This research made use of \textsc{NumPy} \citep{Harris2020}, \textsc{SciPy} \citep{SciPy2020}, \textsc{IPython} \citep{Ipython}. For the plots we used \textsc{Matplotlib}, a Python library for publication quality graphics \citep{Hunter2007}.

\section*{Data Availability}

The data underlying this article are publicly available at the following link \href{https://zenodo.org/record/6418977}{10.5281/zenodo.6418976}.



\bibliographystyle{mnras}
\bibliography{bib} 

\begin{thebibliography}{}
\makeatletter
\relax
\def\mn@urlcharsother{\let\do\@makeother \do\$\do\&\do\#\do\^\do\_\do\%\do\~}
\def\mn@doi{\begingroup\mn@urlcharsother \@ifnextchar [ {\mn@doi@}
  {\mn@doi@[]}}
\def\mn@doi@[#1]#2{\def\@tempa{#1}\ifx\@tempa\@empty \href
  {http://dx.doi.org/#2} {doi:#2}\else \href {http://dx.doi.org/#2} {#1}\fi
  \endgroup}
\def\mn@eprint#1#2{\mn@eprint@#1:#2::\@nil}
\def\mn@eprint@arXiv#1{\href {http://arxiv.org/abs/#1} {{\tt arXiv:#1}}}
\def\mn@eprint@dblp#1{\href {http://dblp.uni-trier.de/rec/bibtex/#1.xml}
  {dblp:#1}}
\def\mn@eprint@#1:#2:#3:#4\@nil{\def\@tempa {#1}\def\@tempb {#2}\def\@tempc
  {#3}\ifx \@tempc \@empty \let \@tempc \@tempb \let \@tempb \@tempa \fi \ifx
  \@tempb \@empty \def\@tempb {arXiv}\fi \@ifundefined
  {mn@eprint@\@tempb}{\@tempb:\@tempc}{\expandafter \expandafter \csname
  mn@eprint@\@tempb\endcsname \expandafter{\@tempc}}}

\bibitem[\protect\citeauthoryear{{Abbott} et~al.,}{{Abbott}
  et~al.}{2020a}]{abbottGW190521}
{Abbott} R.,  et~al., 2020a, \mn@doi [\prl] {10.1103/PhysRevLett.125.101102},
  \href {https://ui.adsabs.harvard.edu/abs/2020PhRvL.125j1102A} {125, 101102}

\bibitem[\protect\citeauthoryear{{Abbott} et~al.,}{{Abbott}
  et~al.}{2020b}]{abbottGW190521astro}
{Abbott} R.,  et~al., 2020b, \mn@doi [\apjl] {10.3847/2041-8213/aba493}, \href
  {https://ui.adsabs.harvard.edu/abs/2020ApJ...900L..13A} {900, L13}

\bibitem[\protect\citeauthoryear{{Abbott} et~al.,}{{Abbott}
  et~al.}{2021a}]{abbottGWTC2.1}
{Abbott} R.,  et~al., 2021a, arXiv e-prints, \href
  {https://ui.adsabs.harvard.edu/abs/2021arXiv210801045T} {p. arXiv:2108.01045}

\bibitem[\protect\citeauthoryear{{Abbott} et~al.,}{{Abbott}
  et~al.}{2021b}]{abbottGWTC3}
{Abbott} R.,  et~al., 2021b, arXiv e-prints, \href
  {https://ui.adsabs.harvard.edu/abs/2021arXiv211103606T} {p. arXiv:2111.03606}

\bibitem[\protect\citeauthoryear{{Ballone}, {Costa}, {Mapelli}  \&
  {MacLeod}}{{Ballone} et~al.}{2022}]{Ballone2022}
{Ballone} A.,  {Costa} G.,  {Mapelli} M.,   {MacLeod} M.,  2022, arXiv
  e-prints, \href {https://ui.adsabs.harvard.edu/abs/2022arXiv220403493B} {p.
  arXiv:2204.03493}

\bibitem[\protect\citeauthoryear{{Belczynski} et~al.,}{{Belczynski}
  et~al.}{2016}]{belczynski2016pair}
{Belczynski} K.,  et~al., 2016, \mn@doi [\aap] {10.1051/0004-6361/201628980},
  \href {https://ui.adsabs.harvard.edu/abs/2016A&A...594A..97B} {594, A97}

\bibitem[\protect\citeauthoryear{{Boekholt}, {Schleicher}, {Fellhauer},
  {Klessen}, {Reinoso}, {Stutz}  \& {Haemmerl{\'e}}}{{Boekholt}
  et~al.}{2018}]{boekholt2018}
{Boekholt} T.~C.~N.,  {Schleicher} D.~R.~G.,  {Fellhauer} M.,  {Klessen} R.~S.,
   {Reinoso} B.,  {Stutz} A.~M.,   {Haemmerl{\'e}} L.,  2018, \mn@doi [\mnras]
  {10.1093/mnras/sty208}, \href
  {https://ui.adsabs.harvard.edu/abs/2018MNRAS.476..366B} {476, 366}

\bibitem[\protect\citeauthoryear{{Bressan}, {Chiosi}  \& {Bertelli}}{{Bressan}
  et~al.}{1981}]{Bressan1981}
{Bressan} A.~G.,  {Chiosi} C.,   {Bertelli} G.,  1981, \aap, \href
  {https://ui.adsabs.harvard.edu/\#abs/1981A&A...102...25B} {102, 25}

\bibitem[\protect\citeauthoryear{{Bressan}, {Marigo}, {Girardi}, {Salasnich},
  {Dal Cero}, {Rubele}  \& {Nanni}}{{Bressan} et~al.}{2012}]{bressan2012}
{Bressan} A.,  {Marigo} P.,  {Girardi} L.,  {Salasnich} B.,  {Dal Cero} C.,
  {Rubele} S.,   {Nanni} A.,  2012, \mn@doi [\mnras]
  {10.1111/j.1365-2966.2012.21948.x}, \href
  {https://ui.adsabs.harvard.edu/abs/2012MNRAS.427..127B} {427, 127}

\bibitem[\protect\citeauthoryear{{Brott} et~al.,}{{Brott}
  et~al.}{2011}]{Brott2011}
{Brott} I.,  et~al., 2011, \mn@doi [\aap] {10.1051/0004-6361/201016113}, 530,
  A115

\bibitem[\protect\citeauthoryear{{Caffau}, {Ludwig}, {Steffen}, {Freytag}  \&
  {Bonifacio}}{{Caffau} et~al.}{2011}]{Caffau2011}
{Caffau} E.,  {Ludwig} H.~G.,  {Steffen} M.,  {Freytag} B.,   {Bonifacio} P.,
  2011, \mn@doi [\solphys] {10.1007/s11207-010-9541-4}, 268, 255

\bibitem[\protect\citeauthoryear{{Chieffi} et~al.,}{{Chieffi}
  et~al.}{2021}]{chieffi2021}
{Chieffi} A.,  et~al., 2021, \mn@doi [\apj] {10.3847/1538-4357/ac06ca}, \href
  {https://ui.adsabs.harvard.edu/abs/2021ApJ...916...79C} {916, 79}

\bibitem[\protect\citeauthoryear{{Costa}, {Girardi}, {Bressan}, {Marigo},
  {Rodrigues}, {Chen}, {Lanza}  \& {Goudfrooij}}{{Costa}
  et~al.}{2019a}]{costa2019a}
{Costa} G.,  {Girardi} L.,  {Bressan} A.,  {Marigo} P.,  {Rodrigues} T.~S.,
  {Chen} Y.,  {Lanza} A.,   {Goudfrooij} P.,  2019a, \mn@doi [\mnras]
  {10.1093/mnras/stz728}, \href
  {https://ui.adsabs.harvard.edu/abs/2019MNRAS.485.4641C} {485, 4641}

\bibitem[\protect\citeauthoryear{{Costa}, {Girardi}, {Bressan}, {Chen},
  {Goudfrooij}, {Marigo}, {Rodrigues}  \& {Lanza}}{{Costa}
  et~al.}{2019b}]{Costa2019b}
{Costa} G.,  {Girardi} L.,  {Bressan} A.,  {Chen} Y.,  {Goudfrooij} P.,
  {Marigo} P.,  {Rodrigues} T.~S.,   {Lanza} A.,  2019b, \mn@doi [\aap]
  {10.1051/0004-6361/201936409}, \href
  {https://ui.adsabs.harvard.edu/abs/2019A&A...631A.128C} {631, A128}

\bibitem[\protect\citeauthoryear{{Costa}, {Bressan}, {Mapelli}, {Marigo},
  {Iorio}  \& {Spera}}{{Costa} et~al.}{2021}]{costa2021}
{Costa} G.,  {Bressan} A.,  {Mapelli} M.,  {Marigo} P.,  {Iorio} G.,   {Spera}
  M.,  2021, \mn@doi [\mnras] {10.1093/mnras/staa3916}, \href
  {https://ui.adsabs.harvard.edu/abs/2021MNRAS.501.4514C} {501, 4514}

\bibitem[\protect\citeauthoryear{{Di Carlo}, {Giacobbo}, {Mapelli}, {Pasquato},
  {Spera}, {Wang}  \& {Haardt}}{{Di Carlo} et~al.}{2019}]{dicarlo2019}
{Di Carlo} U.~N.,  {Giacobbo} N.,  {Mapelli} M.,  {Pasquato} M.,  {Spera} M.,
  {Wang} L.,   {Haardt} F.,  2019, \mn@doi [\mnras] {10.1093/mnras/stz1453},
  \href {https://ui.adsabs.harvard.edu/abs/2019MNRAS.487.2947D} {487, 2947}

\bibitem[\protect\citeauthoryear{{Di Carlo}, {Mapelli}, {Bouffanais},
  {Giacobbo}, {Santoliquido}, {Bressan}, {Spera}  \& {Haardt}}{{Di Carlo}
  et~al.}{2020a}]{dicarlo2020a}
{Di Carlo} U.~N.,  {Mapelli} M.,  {Bouffanais} Y.,  {Giacobbo} N.,
  {Santoliquido} F.,  {Bressan} A.,  {Spera} M.,   {Haardt} F.,  2020a, \mn@doi
  [\mnras] {10.1093/mnras/staa1997}, \href
  {https://ui.adsabs.harvard.edu/abs/2020MNRAS.497.1043D} {497, 1043}

\bibitem[\protect\citeauthoryear{{Di Carlo} et~al.,}{{Di Carlo}
  et~al.}{2020b}]{dicarlo2020b}
{Di Carlo} U.~N.,  et~al., 2020b, \mn@doi [\mnras] {10.1093/mnras/staa2286},
  \href {https://ui.adsabs.harvard.edu/abs/2020MNRAS.498..495D} {498, 495}

\bibitem[\protect\citeauthoryear{{Di Carlo} et~al.,}{{Di Carlo}
  et~al.}{2021}]{dicarlo2021}
{Di Carlo} U.~N.,  et~al., 2021, \mn@doi [\mnras] {10.1093/mnras/stab2390},
  \href {https://ui.adsabs.harvard.edu/abs/2021MNRAS.507.5132D} {507, 5132}

\bibitem[\protect\citeauthoryear{{Farmer}, {Fields}, {Petermann}, {Dessart},
  {Cantiello}, {Paxton}  \& {Timmes}}{{Farmer} et~al.}{2016}]{Farmer2016}
{Farmer} R.,  {Fields} C.~E.,  {Petermann} I.,  {Dessart} L.,  {Cantiello} M.,
  {Paxton} B.,   {Timmes} F.~X.,  2016, \mn@doi [\apjs]
  {10.3847/1538-4365/227/2/22}, \href
  {https://ui.adsabs.harvard.edu/abs/2016ApJS..227...22F} {227, 22}

\bibitem[\protect\citeauthoryear{{Farmer}, {Renzo}, {de Mink}, {Marchant}  \&
  {Justham}}{{Farmer} et~al.}{2019}]{farmer2019}
{Farmer} R.,  {Renzo} M.,  {de Mink} S.~E.,  {Marchant} P.,   {Justham} S.,
  2019, \mn@doi [\apj] {10.3847/1538-4357/ab518b}, \href
  {https://ui.adsabs.harvard.edu/abs/2019ApJ...887...53F} {887, 53}

\bibitem[\protect\citeauthoryear{{Farmer}, {Renzo}, {de Mink}, {Fishbach}  \&
  {Justham}}{{Farmer} et~al.}{2020}]{farmer2020}
{Farmer} R.,  {Renzo} M.,  {de Mink} S.~E.,  {Fishbach} M.,   {Justham} S.,
  2020, \mn@doi [\apjl] {10.3847/2041-8213/abbadd}, \href
  {https://ui.adsabs.harvard.edu/abs/2020ApJ...902L..36F} {902, L36}

\bibitem[\protect\citeauthoryear{{Farrell}, {Groh}, {Hirschi}, {Murphy},
  {Kaiser}, {Ekstr{\"o}m}, {Georgy}  \& {Meynet}}{{Farrell}
  et~al.}{2021}]{farrell2021}
{Farrell} E.,  {Groh} J.~H.,  {Hirschi} R.,  {Murphy} L.,  {Kaiser} E.,
  {Ekstr{\"o}m} S.,  {Georgy} C.,   {Meynet} G.,  2021, \mn@doi [\mnras]
  {10.1093/mnrasl/slaa196}, \href
  {https://ui.adsabs.harvard.edu/abs/2021MNRAS.502L..40F} {502, L40}

\bibitem[\protect\citeauthoryear{{Fern{\'a}ndez}, {Quataert}, {Kashiyama}  \&
  {Coughlin}}{{Fern{\'a}ndez} et~al.}{2018}]{Fernandez2018}
{Fern{\'a}ndez} R.,  {Quataert} E.,  {Kashiyama} K.,   {Coughlin} E.~R.,  2018,
  \mn@doi [\mnras] {10.1093/mnras/sty306}, \href
  {https://ui.adsabs.harvard.edu/abs/2018MNRAS.476.2366F} {476, 2366}

\bibitem[\protect\citeauthoryear{{Ferraro} et~al.,}{{Ferraro}
  et~al.}{2012}]{ferraro2012}
{Ferraro} F.~R.,  et~al., 2012, \mn@doi [\nat] {10.1038/nature11686}, \href
  {https://ui.adsabs.harvard.edu/abs/2012Natur.492..393F} {492, 393}

\bibitem[\protect\citeauthoryear{{Freitag}, {G{\"u}rkan}  \& {Rasio}}{{Freitag}
  et~al.}{2006}]{freitag2006}
{Freitag} M.,  {G{\"u}rkan} M.~A.,   {Rasio} F.~A.,  2006, \mn@doi [\mnras]
  {10.1111/j.1365-2966.2006.10096.x}, \href
  {https://ui.adsabs.harvard.edu/abs/2006MNRAS.368..141F} {368, 141}

\bibitem[\protect\citeauthoryear{{Gaburov}, {Lombardi}, {Portegies Zwart}  \&
  {Rasio}}{{Gaburov} et~al.}{2018}]{gaburov2018}
{Gaburov} E.,  {Lombardi} James~C. J.,  {Portegies Zwart} S.,   {Rasio} F.~A.,
  2018, {StarSmasher: Smoothed Particle Hydrodynamics code for smashing stars
  and planets} (\mn@eprint {ascl} {1805.010})

\bibitem[\protect\citeauthoryear{{Giacobbo}, {Mapelli}  \& {Spera}}{{Giacobbo}
  et~al.}{2018}]{Giacobbo2018}
{Giacobbo} N.,  {Mapelli} M.,   {Spera} M.,  2018, \mn@doi [\mnras]
  {10.1093/mnras/stx2933}, \href
  {https://ui.adsabs.harvard.edu/abs/2018MNRAS.474.2959G} {474, 2959}

\bibitem[\protect\citeauthoryear{{Giersz}, {Leigh}, {Hypki}, {L{\"u}tzgendorf}
  \& {Askar}}{{Giersz} et~al.}{2015}]{giersz2015}
{Giersz} M.,  {Leigh} N.,  {Hypki} A.,  {L{\"u}tzgendorf} N.,   {Askar} A.,
  2015, \mn@doi [\mnras] {10.1093/mnras/stv2162}, \href
  {https://ui.adsabs.harvard.edu/abs/2015MNRAS.454.3150G} {454, 3150}

\bibitem[\protect\citeauthoryear{{Glebbeek}, {Pols}  \& {Hurley}}{{Glebbeek}
  et~al.}{2008}]{glebbeek2008}
{Glebbeek} E.,  {Pols} O.~R.,   {Hurley} J.~R.,  2008, \mn@doi [\aap]
  {10.1051/0004-6361:200809930}, \href
  {https://ui.adsabs.harvard.edu/abs/2008A&A...488.1007G} {488, 1007}

\bibitem[\protect\citeauthoryear{{Glebbeek}, {Gaburov}, {de Mink}, {Pols}  \&
  {Portegies Zwart}}{{Glebbeek} et~al.}{2009}]{glebbeek2009}
{Glebbeek} E.,  {Gaburov} E.,  {de Mink} S.~E.,  {Pols} O.~R.,   {Portegies
  Zwart} S.~F.,  2009, \mn@doi [\aap] {10.1051/0004-6361/200810425}, \href
  {https://ui.adsabs.harvard.edu/abs/2009A&A...497..255G} {497, 255}

\bibitem[\protect\citeauthoryear{{Glebbeek}, {Gaburov}, {Portegies Zwart}  \&
  {Pols}}{{Glebbeek} et~al.}{2013}]{Glebbeek2013}
{Glebbeek} E.,  {Gaburov} E.,  {Portegies Zwart} S.,   {Pols} O.~R.,  2013,
  \mn@doi [\mnras] {10.1093/mnras/stt1268}, \href
  {https://ui.adsabs.harvard.edu/abs/2013MNRAS.434.3497G} {434, 3497}

\bibitem[\protect\citeauthoryear{{G{\"u}rkan}, {Fregeau}  \&
  {Rasio}}{{G{\"u}rkan} et~al.}{2006}]{guerkan2006}
{G{\"u}rkan} M.~A.,  {Fregeau} J.~M.,   {Rasio} F.~A.,  2006, \mn@doi [\apjl]
  {10.1086/503295}, \href
  {https://ui.adsabs.harvard.edu/abs/2006ApJ...640L..39G} {640, L39}

\bibitem[\protect\citeauthoryear{Harris et~al.,}{Harris
  et~al.}{2020}]{Harris2020}
Harris C.~R.,  et~al., 2020, \mn@doi [Nature] {10.1038/s41586-020-2649-2}, 585,
  357

\bibitem[\protect\citeauthoryear{{Heger} \& {Woosley}}{{Heger} \&
  {Woosley}}{2002}]{heger2002}
{Heger} A.,  {Woosley} S.~E.,  2002, \mn@doi [\apj] {10.1086/338487}, \href
  {https://ui.adsabs.harvard.edu/abs/2002ApJ...567..532H} {567, 532}

\bibitem[\protect\citeauthoryear{Hunter}{Hunter}{2007}]{Hunter2007}
Hunter J.~D.,  2007, \mn@doi [Computing In Science \& Engineering]
  {10.1109/MCSE.2007.55}, 9, 90

\bibitem[\protect\citeauthoryear{{Kippenhahn}, {Weigert}  \&
  {Weiss}}{{Kippenhahn} et~al.}{2012}]{Kippenhahn2012}
{Kippenhahn} R.,  {Weigert} A.,   {Weiss} A.,  2012, {Stellar Structure and
  Evolution}.
Stellar Structure and Evolution: , Astronomy and Astrophysics Library. ISBN
  978-3-642-30255-8. Springer-Verlag Berlin Heidelberg, 2012,
  \mn@doi{10.1007/978-3-642-30304-3}

\bibitem[\protect\citeauthoryear{{Kremer} et~al.,}{{Kremer}
  et~al.}{2020}]{kremer2020}
{Kremer} K.,  et~al., 2020, \mn@doi [\apj] {10.3847/1538-4357/abb945}, \href
  {https://ui.adsabs.harvard.edu/abs/2020ApJ...903...45K} {903, 45}

\bibitem[\protect\citeauthoryear{{Kunitomo}, {Guillot}, {Takeuchi}  \&
  {Ida}}{{Kunitomo} et~al.}{2017}]{Kunitomo2017}
{Kunitomo} M.,  {Guillot} T.,  {Takeuchi} T.,   {Ida} S.,  2017, \mn@doi [\aap]
  {10.1051/0004-6361/201628260}, \href
  {https://ui.adsabs.harvard.edu/abs/2017A&A...599A..49K} {599, A49}

\bibitem[\protect\citeauthoryear{{Ledoux}}{{Ledoux}}{1947}]{Ledoux1947}
{Ledoux} P.,  1947, \mn@doi [\apj] {10.1086/144905}, \href
  {https://ui.adsabs.harvard.edu/abs/1947ApJ...105..305L} {105, 305}

\bibitem[\protect\citeauthoryear{{Lovegrove} \& {Woosley}}{{Lovegrove} \&
  {Woosley}}{2013}]{Lovegrove2013}
{Lovegrove} E.,  {Woosley} S.~E.,  2013, \mn@doi [\apj]
  {10.1088/0004-637X/769/2/109}, \href
  {https://ui.adsabs.harvard.edu/abs/2013ApJ...769..109L} {769, 109}

\bibitem[\protect\citeauthoryear{{Lovegrove}, {Woosley}  \&
  {Zhang}}{{Lovegrove} et~al.}{2017}]{Lovegrove2017}
{Lovegrove} E.,  {Woosley} S.~E.,   {Zhang} W.,  2017, \mn@doi [\apj]
  {10.3847/1538-4357/aa7b7d}, \href
  {https://ui.adsabs.harvard.edu/abs/2017ApJ...845..103L} {845, 103}

\bibitem[\protect\citeauthoryear{{Maeder}}{{Maeder}}{1975}]{Maeder1975}
{Maeder} A.,  1975, \aap, \href
  {https://ui.adsabs.harvard.edu/\#abs/1975A&A....40..303M} {40, 303}

\bibitem[\protect\citeauthoryear{{Mapelli}}{{Mapelli}}{2016}]{mapelli2016}
{Mapelli} M.,  2016, \mn@doi [\mnras] {10.1093/mnras/stw869}, \href
  {https://ui.adsabs.harvard.edu/abs/2016MNRAS.459.3432M} {459, 3432}

\bibitem[\protect\citeauthoryear{{Mapelli}, {Sigurdsson}, {Colpi}, {Ferraro},
  {Possenti}, {Rood}, {Sills}  \& {Beccari}}{{Mapelli}
  et~al.}{2004}]{mapelli2004}
{Mapelli} M.,  {Sigurdsson} S.,  {Colpi} M.,  {Ferraro} F.~R.,  {Possenti} A.,
  {Rood} R.~T.,  {Sills} A.,   {Beccari} G.,  2004, \mn@doi [\apjl]
  {10.1086/386370}, \href
  {https://ui.adsabs.harvard.edu/abs/2004ApJ...605L..29M} {605, L29}

\bibitem[\protect\citeauthoryear{{Mapelli}, {Sigurdsson}, {Ferraro}, {Colpi},
  {Possenti}  \& {Lanzoni}}{{Mapelli} et~al.}{2006}]{mapelli2006}
{Mapelli} M.,  {Sigurdsson} S.,  {Ferraro} F.~R.,  {Colpi} M.,  {Possenti} A.,
   {Lanzoni} B.,  2006, \mn@doi [\mnras] {10.1111/j.1365-2966.2006.11038.x},
  \href {https://ui.adsabs.harvard.edu/abs/2006MNRAS.373..361M} {373, 361}

\bibitem[\protect\citeauthoryear{{Mapelli}, {Spera}, {Montanari}, {Limongi},
  {Chieffi}, {Giacobbo}, {Bressan}  \& {Bouffanais}}{{Mapelli}
  et~al.}{2020}]{mapelli2020}
{Mapelli} M.,  {Spera} M.,  {Montanari} E.,  {Limongi} M.,  {Chieffi} A.,
  {Giacobbo} N.,  {Bressan} A.,   {Bouffanais} Y.,  2020, \mn@doi [\apj]
  {10.3847/1538-4357/ab584d}, \href
  {https://ui.adsabs.harvard.edu/abs/2020ApJ...888...76M} {888, 76}

\bibitem[\protect\citeauthoryear{{Marchant} \& {Moriya}}{{Marchant} \&
  {Moriya}}{2020}]{marchant2021}
{Marchant} P.,  {Moriya} T.~J.,  2020, \mn@doi [\aap]
  {10.1051/0004-6361/202038902}, \href
  {https://ui.adsabs.harvard.edu/abs/2020A&A...640L..18M} {640, L18}

\bibitem[\protect\citeauthoryear{{O'Connor} \& {Ott}}{{O'Connor} \&
  {Ott}}{2011}]{OConnor2011}
{O'Connor} E.,  {Ott} C.~D.,  2011, \mn@doi [\apj]
  {10.1088/0004-637X/730/2/70}, \href
  {https://ui.adsabs.harvard.edu/abs/2011ApJ...730...70O} {730, 70}

\bibitem[\protect\citeauthoryear{{Paxton}, {Bildsten}, {Dotter}, {Herwig},
  {Lesaffre}  \& {Timmes}}{{Paxton} et~al.}{2011}]{Paxton2011}
{Paxton} B.,  {Bildsten} L.,  {Dotter} A.,  {Herwig} F.,  {Lesaffre} P.,
  {Timmes} F.,  2011, \mn@doi [\apjs] {10.1088/0067-0049/192/1/3}, \href
  {https://ui.adsabs.harvard.edu/abs/2011ApJS..192....3P} {192, 3}

\bibitem[\protect\citeauthoryear{{Paxton} et~al.,}{{Paxton}
  et~al.}{2013}]{Paxton2013}
{Paxton} B.,  et~al., 2013, \mn@doi [\apjs] {10.1088/0067-0049/208/1/4}, \href
  {https://ui.adsabs.harvard.edu/abs/2013ApJS..208....4P} {208, 4}

\bibitem[\protect\citeauthoryear{{Paxton} et~al.,}{{Paxton}
  et~al.}{2015}]{Paxton2015}
{Paxton} B.,  et~al., 2015, \mn@doi [\apjs] {10.1088/0067-0049/220/1/15}, \href
  {https://ui.adsabs.harvard.edu/abs/2015ApJS..220...15P} {220, 15}

\bibitem[\protect\citeauthoryear{{Paxton} et~al.,}{{Paxton}
  et~al.}{2018}]{Paxton2018}
{Paxton} B.,  et~al., 2018, \mn@doi [\apjs] {10.3847/1538-4365/aaa5a8}, \href
  {https://ui.adsabs.harvard.edu/abs/2018ApJS..234...34P} {234, 34}

\bibitem[\protect\citeauthoryear{{Paxton} et~al.,}{{Paxton}
  et~al.}{2019}]{Paxton2019}
{Paxton} B.,  et~al., 2019, \mn@doi [\apjs] {10.3847/1538-4365/ab2241}, \href
  {https://ui.adsabs.harvard.edu/abs/2019ApJS..243...10P} {243, 10}

\bibitem[\protect\citeauthoryear{{Perez} \& {Granger}}{{Perez} \&
  {Granger}}{2007}]{Ipython}
{Perez} F.,  {Granger} B.~E.,  2007, \mn@doi [Computing in Science Engineering]
  {10.1109/MCSE.2007.53}, 9, 21

\bibitem[\protect\citeauthoryear{{Portegies Zwart}}{{Portegies
  Zwart}}{2019}]{portegieszwart2018}
{Portegies Zwart} S.,  2019, \mn@doi [\aap] {10.1051/0004-6361/201833485},
  \href {https://ui.adsabs.harvard.edu/abs/2019A&A...621L..10P} {621, L10}

\bibitem[\protect\citeauthoryear{{Portegies Zwart} \& {McMillan}}{{Portegies
  Zwart} \& {McMillan}}{2002}]{portegieszwart2002}
{Portegies Zwart} S.~F.,  {McMillan} S. L.~W.,  2002, \mn@doi [\apj]
  {10.1086/341798}, \href
  {https://ui.adsabs.harvard.edu/abs/2002ApJ...576..899P} {576, 899}

\bibitem[\protect\citeauthoryear{{Portegies Zwart}, {Hut}, {McMillan}  \&
  {Verbunt}}{{Portegies Zwart} et~al.}{1997}]{portegieszwart1997}
{Portegies Zwart} S.~F.,  {Hut} P.,  {McMillan} S. L.~W.,   {Verbunt} F.,
  1997, \aap, \href {https://ui.adsabs.harvard.edu/abs/1997A&A...328..143P}
  {328, 143}

\bibitem[\protect\citeauthoryear{{Portegies Zwart}, {Makino}, {McMillan}  \&
  {Hut}}{{Portegies Zwart} et~al.}{1999}]{portegieszwart1999}
{Portegies Zwart} S.~F.,  {Makino} J.,  {McMillan} S.~L.~W.,   {Hut} P.,  1999,
  \aap, \href {https://ui.adsabs.harvard.edu/abs/1999A&A...348..117P} {348,
  117}

\bibitem[\protect\citeauthoryear{{Portegies Zwart}, {Baumgardt}, {Hut},
  {Makino}  \& {McMillan}}{{Portegies Zwart} et~al.}{2004}]{portegieszwart2004}
{Portegies Zwart} S.~F.,  {Baumgardt} H.,  {Hut} P.,  {Makino} J.,   {McMillan}
  S. L.~W.,  2004, \mn@doi [\nat] {10.1038/nature02448}, \href
  {https://ui.adsabs.harvard.edu/abs/2004Natur.428..724P} {428, 724}

\bibitem[\protect\citeauthoryear{{Powell}, {M{\"u}ller}  \& {Heger}}{{Powell}
  et~al.}{2021}]{Powell2021}
{Powell} J.,  {M{\"u}ller} B.,   {Heger} A.,  2021, \mn@doi [\mnras]
  {10.1093/mnras/stab614}, \href
  {https://ui.adsabs.harvard.edu/abs/2021MNRAS.503.2108P} {503, 2108}

\bibitem[\protect\citeauthoryear{{Rahman}, {Janka}, {Stockinger}  \&
  {Woosley}}{{Rahman} et~al.}{2022}]{Rahman2022}
{Rahman} N.,  {Janka} H.~T.,  {Stockinger} G.,   {Woosley} S.~E.,  2022,
  \mn@doi [\mnras] {10.1093/mnras/stac758}, \href
  {https://ui.adsabs.harvard.edu/abs/2022MNRAS.512.4503R} {512, 4503}

\bibitem[\protect\citeauthoryear{{Renzo}, {Farmer}, {Justham}, {G{\"o}tberg},
  {de Mink}, {Zapartas}, {Marchant}  \& {Smith}}{{Renzo}
  et~al.}{2020a}]{Renzo2020a}
{Renzo} M.,  {Farmer} R.,  {Justham} S.,  {G{\"o}tberg} Y.,  {de Mink} S.~E.,
  {Zapartas} E.,  {Marchant} P.,   {Smith} N.,  2020a, \mn@doi [\aap]
  {10.1051/0004-6361/202037710}, \href
  {https://ui.adsabs.harvard.edu/abs/2020A&A...640A..56R} {640, A56}

\bibitem[\protect\citeauthoryear{{Renzo}, {Cantiello}, {Metzger}  \&
  {Jiang}}{{Renzo} et~al.}{2020b}]{renzo2020}
{Renzo} M.,  {Cantiello} M.,  {Metzger} B.~D.,   {Jiang} Y.~F.,  2020b, \mn@doi
  [\apjl] {10.3847/2041-8213/abc6a6}, \href
  {https://ui.adsabs.harvard.edu/abs/2020ApJ...904L..13R} {904, L13}

\bibitem[\protect\citeauthoryear{{Rizzuto} et~al.,}{{Rizzuto}
  et~al.}{2021}]{rizzuto2021}
{Rizzuto} F.~P.,  et~al., 2021, \mn@doi [\mnras] {10.1093/mnras/staa3634},
  \href {https://ui.adsabs.harvard.edu/abs/2021MNRAS.501.5257R} {501, 5257}

\bibitem[\protect\citeauthoryear{{Schwarzschild}}{{Schwarzschild}}{1958}]{Schwarzschild1958}
{Schwarzschild} M.,  1958, {Structure and evolution of the stars.}.
Princeton, Princeton University Press, 1958.

\bibitem[\protect\citeauthoryear{{Sigurdsson}, {Davies}  \&
  {Bolte}}{{Sigurdsson} et~al.}{1994}]{sigurdsson1994}
{Sigurdsson} S.,  {Davies} M.~B.,   {Bolte} M.,  1994, \mn@doi [\apjl]
  {10.1086/187486}, \href
  {https://ui.adsabs.harvard.edu/abs/1994ApJ...431L.115S} {431, L115}

\bibitem[\protect\citeauthoryear{{Sills}, {Lombardi}, {Bailyn}, {Demarque},
  {Rasio}  \& {Shapiro}}{{Sills} et~al.}{1997}]{sills1997}
{Sills} A.,  {Lombardi} James~C. J.,  {Bailyn} C.~D.,  {Demarque} P.,  {Rasio}
  F.~A.,   {Shapiro} S.~L.,  1997, \mn@doi [\apj] {10.1086/304588}, \href
  {https://ui.adsabs.harvard.edu/abs/1997ApJ...487..290S} {487, 290}

\bibitem[\protect\citeauthoryear{{Sills}, {Faber}, {Lombardi}, {Rasio}  \&
  {Warren}}{{Sills} et~al.}{2001}]{sills2001}
{Sills} A.,  {Faber} J.~A.,  {Lombardi} James~C. J.,  {Rasio} F.~A.,   {Warren}
  A.~R.,  2001, \mn@doi [\apj] {10.1086/318689}, \href
  {https://ui.adsabs.harvard.edu/abs/2001ApJ...548..323S} {548, 323}

\bibitem[\protect\citeauthoryear{{Spera} \& {Mapelli}}{{Spera} \&
  {Mapelli}}{2017}]{spera2017}
{Spera} M.,  {Mapelli} M.,  2017, \mn@doi [\mnras] {10.1093/mnras/stx1576},
  \href {https://ui.adsabs.harvard.edu/abs/2017MNRAS.470.4739S} {470, 4739}

\bibitem[\protect\citeauthoryear{{Spera}, {Mapelli}, {Giacobbo}, {Trani},
  {Bressan}  \& {Costa}}{{Spera} et~al.}{2019}]{spera2019}
{Spera} M.,  {Mapelli} M.,  {Giacobbo} N.,  {Trani} A.~A.,  {Bressan} A.,
  {Costa} G.,  2019, \mn@doi [\mnras] {10.1093/mnras/stz359}, \href
  {https://ui.adsabs.harvard.edu/abs/2019MNRAS.485..889S} {485, 889}

\bibitem[\protect\citeauthoryear{{Stevenson}, {Sampson}, {Powell},
  {Vigna-G{\'o}mez}, {Neijssel}, {Sz{\'e}csi}  \& {Mandel}}{{Stevenson}
  et~al.}{2019}]{stevenson2019}
{Stevenson} S.,  {Sampson} M.,  {Powell} J.,  {Vigna-G{\'o}mez} A.,  {Neijssel}
  C.~J.,  {Sz{\'e}csi} D.,   {Mandel} I.,  2019, \mn@doi [\apj]
  {10.3847/1538-4357/ab3981}, \href
  {https://ui.adsabs.harvard.edu/abs/2019ApJ...882..121S} {882, 121}

\bibitem[\protect\citeauthoryear{{Torniamenti}, {Rastello}, {Mapelli}, {Di
  Carlo}, {Ballone}  \& {Pasquato}}{{Torniamenti}
  et~al.}{2022}]{torniamenti2022}
{Torniamenti} S.,  {Rastello} S.,  {Mapelli} M.,  {Di Carlo} U.~N.,  {Ballone}
  A.,   {Pasquato} M.,  2022, arXiv e-prints, \href
  {https://ui.adsabs.harvard.edu/abs/2022arXiv220308163T} {p. arXiv:2203.08163}

\bibitem[\protect\citeauthoryear{{Vigna-G{\'o}mez}, {Justham}, {Mandel}, {de
  Mink}  \& {Podsiadlowski}}{{Vigna-G{\'o}mez} et~al.}{2019}]{Vigna2019}
{Vigna-G{\'o}mez} A.,  {Justham} S.,  {Mandel} I.,  {de Mink} S.~E.,
  {Podsiadlowski} P.,  2019, \mn@doi [\apjl] {10.3847/2041-8213/ab1bdf}, \href
  {https://ui.adsabs.harvard.edu/abs/2019ApJ...876L..29V} {876, L29}

\bibitem[\protect\citeauthoryear{{Vink}, {Higgins}, {Sander}  \&
  {Sabhahit}}{{Vink} et~al.}{2021}]{vink2021}
{Vink} J.~S.,  {Higgins} E.~R.,  {Sander} A. A.~C.,   {Sabhahit} G.~N.,  2021,
  \mn@doi [\mnras] {10.1093/mnras/stab842}, \href
  {https://ui.adsabs.harvard.edu/abs/2021MNRAS.504..146V} {504, 146}

\bibitem[\protect\citeauthoryear{Virtanen et~al.,}{Virtanen
  et~al.}{2020}]{SciPy2020}
Virtanen P.,  et~al., 2020, \mn@doi [Nature Methods]
  {10.1038/s41592-019-0686-2}, \href {https://rdcu.be/b08Wh} {17, 261}

\bibitem[\protect\citeauthoryear{{Woosley}}{{Woosley}}{2017}]{woosley2017}
{Woosley} S.~E.,  2017, \mn@doi [\apj] {10.3847/1538-4357/836/2/244}, \href
  {https://ui.adsabs.harvard.edu/abs/2017ApJ...836..244W} {836, 244}

\bibitem[\protect\citeauthoryear{{Woosley}}{{Woosley}}{2019}]{woosley2019}
{Woosley} S.~E.,  2019, \mn@doi [\apj] {10.3847/1538-4357/ab1b41}, \href
  {https://ui.adsabs.harvard.edu/abs/2019ApJ...878...49W} {878, 49}

\makeatother
\end{thebibliography}




\appendix

\section{Accretion process heat injection}
\label{a:accretion}

Following \citet{Kunitomo2017} and references therein, we use the parameterization of the heat injected by the accreting material as follows:
\begin{equation}
    L_\mathrm{add} = \frac{\xi_{\rm acc} \,{}G\,{} M \dot{M}}{R} 
	\label{eq:Lacc}
\end{equation}
where $\xi_{\rm acc}$ ranges from 0 to 1, and $M$ and $R$ are the mass and radius of the star, respectively. $\dot{M}$ is the accretion rate in \Msun yr$^{-1}$. We tested different values for the $\xi_{\rm acc}$ parameter (0.1, 0.5, 1.0). In all cases, we find a similar evolution during the accretion, which causes the star to expand, become a \ac{RSG}, and reach the same location in the \ac{HR} diagram.
The injected energy is distributed instantaneously inside the star with a simple model, in which the energy is deposited only in an outer layer of the star of mass $m_\mathrm{ke}$ = 0.5 in mass fraction. The energy deposited per unit of mass is:
\begin{equation}
    \epsilon_\mathrm{add} = \left(\frac{L_\mathrm{add}}{M}\right)\,{} \max{ \left[0, \frac{2}{m_\mathrm{ke}^2} \left(\frac{M_r}{M} - 1 + m_\mathrm{ke}\right)\right]}
	\label{eq:eps}
\end{equation}
where $M_r$ is the mass coordinate.
Finally, we include this additional energy contribution in the structure equation of energy conservation, which reads:
\begin{equation}
    \frac{\partial L}{\partial M_r} = \epsilon_\mathrm{nuc} - \left(T\,{} \frac{\partial S}{\partial t}\right)_{M_r} + \epsilon_\mathrm{add},
	\label{eq:energy}
\end{equation}
where $L$ is the luminosity, $T$ is the temperature, $\epsilon_\mathrm{nuc}$ is the rate of energy production of nuclear reactions, $S$ is the specific entropy of the shell, and $t$ is the time.

To build the post-collision \parsec\ tracks \textsc{post-prim} and \textsc{post-sph}, we adopt the following accretion rate prescription:
\begin{equation}
    \dot{M} (t) = 10^{-4} \times [(t - T_{\rm start})/dt]^{2},
	\label{eq:acc2}
\end{equation}
in which $T_{\rm start}$ is the age of the star when the accretion episode begins, $t$ is the stellar age, and $dt$ is the time-step. During accretion, we use a fixed time-step $dt\sim 0.1$ years.
To avoid numerical convergence issues, we impose a limit on the maximum accretion rate as follows:
\begin{equation}
    \dot{M} (t) = \min{\left[\dot{M}(t), 5~{\rm M}_\odot/{\rm yr}\right]}
 	\label{eq:acc_limit}
\end{equation}
We stop the accretion when \M{coll} is reached.
With the above configuration, the accretion process lasts for about 23 years.





\bsp	
\label{lastpage}
\end{document}